\documentclass[preprint, showpacs,preprintnumbers,amsmath,amssymb,nofootinbib]{revtex4}
\usepackage{mathrsfs}
\usepackage{amsmath}
\usepackage{amsfonts}
\usepackage{latexsym}
\usepackage{amsfonts}
\usepackage{graphicx}
\usepackage{epsf}
\usepackage{dcolumn}
\usepackage{bm}

\newcommand{\bea}[1]{\begin{eqnarray}\label{#1}}
\newcommand{\eea}{\end{eqnarray}}

\def\gsim{ \lower .75ex \hbox{$\sim$} \llap{\raise .27ex \hbox{$>$}} }
\def\lsim{ \lower .75ex \hbox{$\sim$} \llap{\raise .27ex \hbox{$<$}} }

\begin{document}
 \title{Emergent universe from the Ho\v{r}ava-Lifshitz gravity}

\author{Puxun Wu $^{1,2,3}$ and  Hongwei Yu $^{2,1,3}$ \footnote{Corresponding author}}
\address
{$^1$ Center of Non-linear Science and Department of Physics, Ningbo University,  No.818 Fenghua Road, Ningbo, Zhejiang, 315211
 China\\
$^2$ Department of Physics and Institute of  Physics, Hunan Normal
University, Changsha, Hunan 410081, China \\
$^3$ Key Laboratory of Low Dimensional Quantum Structures and Quantum
Control of Ministry of Education, Hunan Normal University, Changsha,
Hunan 410081, China }

\begin{abstract}
We study  the stability of the Einstein static universe in the
Ho\v{r}ava-Lifshitz  (HL) gravity and a generalized version of it
formulated by Sotiriou, Visser and Weifurtner. We find that,
for the HL cosmology, there exists a stable Einstein static state
if the cosmological constant $\Lambda$ is negative. The universe can
stay at this stable state eternally and thus  the big bang
singularity can be avoided. However, in this case, the Universe can
not exit to an inflationary era. For the  Sotiriou, Visser and Weifurtner HL cosmology, if the
cosmic scale factor satisfies certain conditions initially, the
Universe can stay at the stable state past eternally and may undergo
a series of infinite, nonsingular oscillations. Once the parameter
of the equation of state $w$ approaches a critical value,  the
stable critical point coincides with the unstable one, and the
Universe enters an inflationary era. Therefore, the big bang
singularity can be avoided and a subsequent inflation can occur
naturally.

\end{abstract}

 \maketitle

\section{Introduction}\label{sec1}
In the standard cosmological model, the existence of a big bang
singularity in the early universe is still an open problem.  In
order to resolve this problem, Ellis et al. proposed, in the context
of general relativity, a scenario, called an emergent
universe~\cite{Ellis2004a, Ellis2004b}. In this scenario, the space
curvature is positive and the Universe stays, past eternally, in an
Einstein static state  and then evolves to a subsequent inflationary
phase. So, the Universe originates from an Einstein static state,
rather than from a big bang singularity. It is also worth noting
that an Einstein static state as the initial state of the  Universe
is also favored by the entropy considerations~\cite{Gibbons1988}.
However, the Einstein static universe in the classical general
relativity is unstable, which means that it is extremely difficult
for  the Universe  to remain in such an initial static state in a
long time due to the existence of perturbations, such as the quantum
fluctuations. Therefore,  the original emergent model does not seem
to resolve the big bang singularity problem successfully as
expected.

Since in the early epoch, the Universe is presumably under extreme
physical conditions, new effects, such as those stemming from
quantization of gravity, or a modification of general relativity or
even other new physics, may become important.
 As a result, the stability of the Einstein
static state has been examined in various cases~\cite{Carneiro2009,
Mulryne2005, Parisi2007, Wu2009, Lidsey2006, Bohmer2007, Seahra2009,
Bohmer2009, Barrow2003, Barrow2009, Clifton2005,Boehmer2010,
Odrzywolek2009}.  For example, the emergent scenario within the
frameworks of loop quantum gravity and braneworld cosmology have
been discussed in Refs.~\cite{Mulryne2005, Parisi2007, Wu2009,
Lidsey2006},  where it was found that a successful model can be
obtained, while  the stability in the presence of vacuum energy
corresponding to conformally invariant fields has been studied and a
nonsingular
emergent cosmological model was reconstructed~\cite{Carneiro2009}. 
In $f(R)$ gravity,  it was  found that the Einstein static state is
stable under  homogeneous perturbations~\cite{Bohmer2007}, but this
stability is broken by adding inhomogeneous
perturbations~\cite{Seahra2009}. In $f(G)$ gravity, the stability of
the Einstein static state against homogeneous perturbations has been
analyzed in Ref.~\cite{Bohmer2009}, where $G$ is the Guass-Bonnet
term. In addition, Barrow et al.~\cite{Barrow2003} found, with the
covariant techniques, that the Einstein static state is stable for
small inhomogeneous vector and tensor perturbations, as well as for
adiabatic scalar density inhomogeneities with
$c_s^2>0.2$. %

Recently, motivated by  the Lifshitz theory in solid state physics,
Ho\v{r}ava proposed a power-counting renormalizable theory of
gravity, called Ho\v{r}ava-Lifshitz (HL) gravity~\cite{Horava2009}.
In the ultraviolet (UV) limit, HL has  a Lifshitz-like anisotropic
scaling between space and time characterized by the dynamical
critical exponent $z = 3$ and thus breaks the Lorentz invariance,
while in the infrared (IR), it flows to $z=1$. So, it is expected to
reduce to the classical general relativity gravity theory in the low
energy limit. Applying the HL gravity to cosmology,  it has been
found, in a nonflat universe, that the Friedmann equation is
modified by a $\frac{1}{a^4}$ term~\cite{Lu2009, Kiritsis2009,
Calcagni2009}, where $a$ is the scale factor. The cosmological
perturbations with the HL gravity were studied in
Refs.~\cite{Mukohyama2009, Gao2009, Cai2009, Chen2009, Gao20092,
Yamamoto2009, Wang2009, Kobayashi2009, Piao2009,CaiYi2009}, and the
results showed that a scale invariant superhorizon curvature
perturbation could be produced without  inflation. In the original
HL gravity, Ho\v{r}ava assumed two conditions: detailed balance and
projectability. More recently, Sotiriou, Visser and Weifurtner
(SVW)~\cite{Sotiriou2009} suggested to build a general HL theory
with projectability but without detailed-balance conditions. For a
spatially curved Friedmann-Robertson-Walker universe, the SVW generalization gives an extra
$\frac{1}{a^6}$ correction term   and modifies the coefficient of
the $\frac{1}{a^4}$ term in the Friedmann equation as compared to
the HL theory. Let us note here that some other issues in HL gravity
have been dealt with in Refs.~\cite{Lu20092, Lu20093, Lu20094,
Lu20095}.

In this paper, we will discuss the stability of the Einstein static
universe in the contexts  of HL gravity and SVW HL theory,
respectively.
 In the following
section, we briefly review the HL and SVW HL cosmology. In section
III, we analyze the Einstein static solutions and discuss the
stability of these solutions. Finally, in section IV, we present our
main conclusions.

\section{The Ho\v{r}ava-Lifshitz cosmology}
In HL gravity, it is convenient to use the Arnowitt-Deser-Misner decomposition of the
metric
 \begin{eqnarray} ds^2=-N^2
dt^2+g_{ij}(dx^i+N^idt)(dx^j+N^jdt)\;,
\end{eqnarray}
where $g_{ij}$ is the $3$-dimensional spatial metric, $N$ is the
lapse function, $N^i$ is the shift vector, and the coordinates scale
as $t\rightarrow \ell^z t$, $x^i\rightarrow \ell x^i$. In this
paper, we only consider the  $z=3$ case.  Next, we first turn our
attention to the implications of HL gravity in cosmology.

\subsection{The HL cosmology}

The action of  HL gravity consists  of  kinetic and potential terms.
The former is given by
\begin{eqnarray} S_k=\frac{2}{\kappa^2}\int dt
d^3x\sqrt{g}N(K_{ij}K^{ij}-\lambda K^2)\;,\end{eqnarray} where
$K_{ij}=\frac{1}{2N}(\dot{g}_{ij}-\nabla_iN_j-\nabla_jN_i)$ is the
extrinsic curvature, $K=g^{ij}K_{ij}$, $K^{ij}=g^{ik}g^{jl}K_{kl}$,
 and $\lambda$ is a dimensionless parameter. When $\lambda=1$, one
 recovers the kinetic part of the 4-dimensional Einstein-Hilbert action.
 With the detailed-balance condition, the potential term has
the form \begin{eqnarray}S_V=\int dt d^3x\sqrt{g}N\big(\beta
C_{ij}C^{ij}+\gamma\frac{\epsilon^{ijk}}{\sqrt{g}}R_{il}\nabla_jR^l_k+\zeta
R_{ij}R^{ij}+\eta R^2+\xi R+\sigma\big)\;.\end{eqnarray} Here
$R_{ij}$ is the three-dimensional spatial curvature tensor,
$R=g^{ij}R_{ij}$, $\epsilon^{ijk}$ is the antisymmetric tensor with
$\epsilon^{123}=1$ and
$C^{ij}=\frac{\epsilon^{ilk}}{\sqrt{g}}\nabla_k\big(R^j_l-\frac{1}{4}\delta_l^jR\big)$
is the Cotton tensor. The constants $\beta$, $\gamma$, $\zeta$,
$\eta$, $\xi$ and $\sigma$ are defined, respectively, as
\begin{eqnarray}
\beta=-\frac{\kappa^2}{2\omega^4}, \quad \gamma=\frac{\kappa^2 \mu
}{2\omega^2},\quad \zeta=-\frac{\kappa^2 \mu^2 }{8}, \quad
\eta=\frac{\kappa^2 \mu^2 (1-4\lambda) }{32 (1-3\lambda)},\nonumber\\
\quad \xi=\frac{\kappa^2 \mu^2 }{8(1-3\lambda)}\Lambda, \quad
\sigma=-\frac{3\kappa^2 \mu^2
}{8(1-3\lambda)}\Lambda^2\;,\end{eqnarray} where  $\Lambda$ is the
cosmological constant, and $\mu$ and $\omega$ are two coupling
constants. In this case, the emergent speed of light becomes
\begin{eqnarray}\label{cq}
c=\frac{\kappa^2 \mu}{4}\sqrt{\frac{\Lambda}{1-3\lambda}}\;.
\end{eqnarray}
So, for the case where $3\lambda-1>0$,  a negative $\Lambda$ is
required in order to guarantee that the speed of light is real. Let
us note that a positive $\Lambda$ can be obtained by making an
analytical continuation for parameters $\mu$ and $\omega^2$ by
$\mu\rightarrow i\mu$ and $\omega^2
\rightarrow-i\omega^2$~\cite{Lu2009}. In addition, it was found in
Ref.~\cite{Setare2009} that a negative cosmological constant may
disappear in the different geometries, plane symmetric spacetimes,
for example.

For a homogeneous and isotropic universe described by the metric
\begin{eqnarray}
ds^2=-dt^2+a^2(t)\bigg(\frac{dr^2}{1-kr^2}+r^2d^2\Omega\bigg)\;,\end{eqnarray}
 the Friedmann equation in HL gravity can be expressed as
\begin{eqnarray}\frac{6}{\kappa^2}(3\lambda-1)
H^2=\rho-\sigma-\frac{6k\xi}{a^2}-\frac{12k^2(\zeta+3\eta)}{a^4}\;.\end{eqnarray}
Here $k=0, \pm 1$, and $\rho$ is the energy density of a perfect
fluid in the universe, which satisfies the conservation equation
\begin{eqnarray}\dot{\rho}+3H\rho(1+w)=0\;,\end{eqnarray}
where $w=p/\rho$ is the equation of state. In the present paper, a
constant $w$ is considered,  which is a good approximation,  since,
as shown in Refs.~\cite{Ellis2004a, Ellis2004b},  a plateau
potential is required in the past-asymptotic limit in the emergent
scenario.  It is easy to see that, for a spatially flat universe,
this Friedmann equation is the same as that in general relativity.

Now, we define two new constants
\begin{eqnarray}\label{alpha} \beta=\frac{\kappa^2}{6(3\lambda-1)},
\qquad
\alpha=\frac{\mu^2\kappa^4}{48(3\lambda-1)(1-3\lambda)}\;.\end{eqnarray}
Apparently,  $\beta$ is positive if $3\lambda-1>0$ and negative if
$3\lambda-1<0$, whereas $\alpha$ is always negative if $\mu^2>0$.
Using these newly defined constants, the above Friedmann equation
for a closed universe can be reexpressed as
\begin{eqnarray}\label{H2}H^2=\beta \rho+ 3 \alpha
\Lambda^2-\frac{6\alpha\Lambda}{a^2}+\frac{3\alpha}{a^4}\;.\end{eqnarray}
which be further written as
\begin{eqnarray}\label{H21}H^2=\beta \rho+ 3 \alpha
\Lambda^2\bigg[1-\frac{2}{\Lambda{a}^2}+\frac{1}{\Lambda^2{a}^4}\bigg]\;.\end{eqnarray}
Differentiating this equation with time and using the energy
conservation equation, we have
\begin{eqnarray}\label{ddota}\frac{\ddot{ {a}}}{ {a}}=-\frac{1+3w}{2}H^2+
\alpha \Lambda^2\bigg[\frac{9}{2}(1+w)-3
(1+3w)\frac{1}{\Lambda{a}^2}-
\frac{3(1-3w)}{2}\frac{1}{\Lambda^2{a}^4}\bigg]\;.\end{eqnarray}

\subsection{The SVW HL cosmology} Sotiriou, Visser and Weifurtner
generalized the original HL theory by keeping the projectability but
abandoning the detailed-balance conditions~\cite{Sotiriou2009}. In
this case, the modified Friedmann equation has the following form
\begin{eqnarray}\label{SVWEq}
\bigg(1-\frac{3}{2}\lambda
\bigg)H^2=\rho+\frac{\Lambda}{3}-\frac{1}{a^2}+\frac{2\beta_1}{a^4}+\frac{4\beta_2}{a^6}\;,
\end{eqnarray}
where $\lambda$, $\beta_1$ and $\beta_2$ are coupling constants.
Comparing the above equation with that in HL theory, one can see
that the SVW generalization not only  gives  an extra correction
term, but also modifies the coefficients of other terms. It is
interesting to note that the ${1/a^6}$ correction term may also
result from  a radiation fluid within the UV
regime~\cite{Minamitsuji2010}. So, the influence of  such a
radiation fluid within the UV regime on the stability of an Einstein
static state can be regarded as a specific case of the analysis to
be carried out next. Defining two dimensionless constants,
$\bar{\beta}_1=\beta_1 \Lambda$ and $\bar{\beta}_2=\beta_2
\Lambda^2$, we find that the modified Friedmann equation in SVW HL
theory becomes
\begin{eqnarray}\label{SVW2}
\bigg(1-\frac{3}{2}\lambda
\bigg)H^2=\rho+\Lambda\bigg[\frac{1}{3}-\frac{1}{\Lambda{a}^2}+\frac{2\bar{\beta}_1}{\Lambda^2{a}^4}
+\frac{4\bar{\beta}_2}{\Lambda^3{a}^6}\bigg]\;.
\end{eqnarray}
Then, differentiating Eq.~(\ref{SVW2}) with  time, one has
\begin{eqnarray}\label{SVW3}
2\bigg(1-\frac{3}{2}\lambda \bigg)\frac{\ddot{ {a}}}{
{a}}&=&-(1-\frac{3}{2}\lambda)(1+3w)H^2\nonumber
\\ &+&\Lambda
\bigg[(1+w)-\frac{1+3w}{\Lambda{a}^2}+\frac{6w-2}{\Lambda^2{a}^4}\bar{\beta}_1+\frac{12w-12}{\Lambda^3{a}^6}\bar{\beta}_2\bigg]\;.
\end{eqnarray}

\section{The Einstein static solution}
The Einstein static solution is given by the conditions $\dot{a}=0$
and $\ddot{a}=0$,  which imply \begin{eqnarray}a=a_{Es},\qquad
H(a_{Es})=0\;. \end{eqnarray}
\subsection{The HL cosmology}
 From Eq.~(\ref{ddota}), it is easy to that the Einstein static
 solution satisfies the following equation
 \begin{eqnarray}
 \frac{9}{2}(1+w)-3 (1+3w)\frac{1}{\Lambda{a}^2}-
\frac{3(1-3w)}{2}\frac{1}{\Lambda^2{a}^4}=0\;.
\end{eqnarray}
Solving this equation, one obtains two critical points
\begin{eqnarray}\label{S1}Point\; A: \qquad
\frac{1}{{a}^2_{Es}}=\Lambda
\end{eqnarray} and
\begin{eqnarray}\label{S2}Point\; B: \qquad \frac{1}{{a}^2_{Es}}=-\frac{3(1+w)}{1-3w}\Lambda\;. \end{eqnarray}
Substituting these  critical points into  Eq.~(\ref{H2}) reveals
that Point $A$ corresponds to
\begin{eqnarray}\rho_A=0
\end{eqnarray}
and Point $B$  to
\begin{eqnarray}\rho_B=-\frac{48
\alpha}{\beta (1-3w)^2} \Lambda^2\;.
\end{eqnarray}
If $\Lambda$ is negative, Point A is physically meaningless since
$a_{Es}^2=\frac{1}{\Lambda}$ is negative. The existence condition
for Point B is $-\frac{3(1+w)}{1-3w}<0$, which leads to
$-1<w<\frac{1}{3}$. For a positive $\Lambda$ obtained from an
analytical continuation for parameters $\mu$ and $\omega^2$ by
$\mu\rightarrow i\mu$ and $\omega^2 \rightarrow-i\omega^2$, which
yields a positive $\alpha$  since  $\mu^2$ becomes negative after
the analytical continuation [refer to Eq.~(\ref{alpha})], it seems
that Point A exists and so does Point B if $w$ satisfies the
condition $w<-1$ or $w>\frac{1}{3}$. However, in this case,  the
energy density corresponding to Point B, $\rho_B$, becomes a
negative since $\alpha$ is positive, which is meaningless. So,  in
the case of a positive $\Lambda$, only point A exists physically.

In order to study the stability of these critical points, we
introduce two variables \begin{eqnarray} x_1=a, \quad x_2=\dot{a}\;.
\end{eqnarray}
They obey the following equations
\begin{eqnarray} \dot{x}_1=  x_2\;,
\end{eqnarray}
\begin{eqnarray} \dot{x}_2= -\frac{1+3w}{2}\frac{ x^2_2}{x_1}+\frac{9}{2}\alpha \Lambda^2(1+w)x_1-3\alpha\Lambda (1+3w)\frac{1}{x_1}-
\frac{3\alpha(1-3w)}{2}\frac{1}{x_1^3}\;.
\end{eqnarray}
Linearizing the system described  by the above two equations near
critical points, one can obtain  the eigenvalues of  the coefficient
matrix, which determine the stability of these critical points.
After some calculations, we get the eigenvalue $\vartheta^2$:
 \begin{eqnarray}\label{eigen1}
Point \; A: \qquad \vartheta^2=12 \alpha \Lambda^2 \;,\end{eqnarray}
\begin{eqnarray}\label{eigen2}
Point \; B: \qquad \vartheta^2=-12 \alpha\Lambda \frac{1}{
{a}_{Es}^2}\;.\end{eqnarray} If $\vartheta^2<0$, the corresponding
equilibrium point is a center point, otherwise it is a saddle one. In
order to analyze the stability of the critical points in detail,  we
now divide our discussions into two cases, i.e., $\Lambda<0$ and
$\Lambda>0$.

\subsubsection{ $\Lambda<0$ }
\begin{figure}[htbp]
\includegraphics[width=5cm]{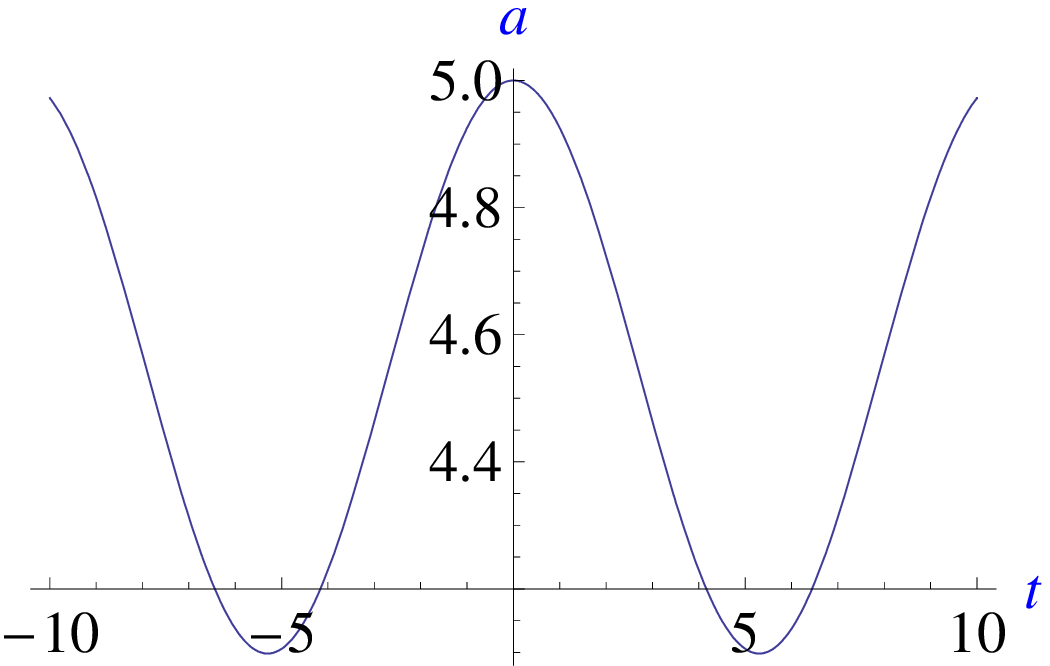}\quad\includegraphics[width=5cm]{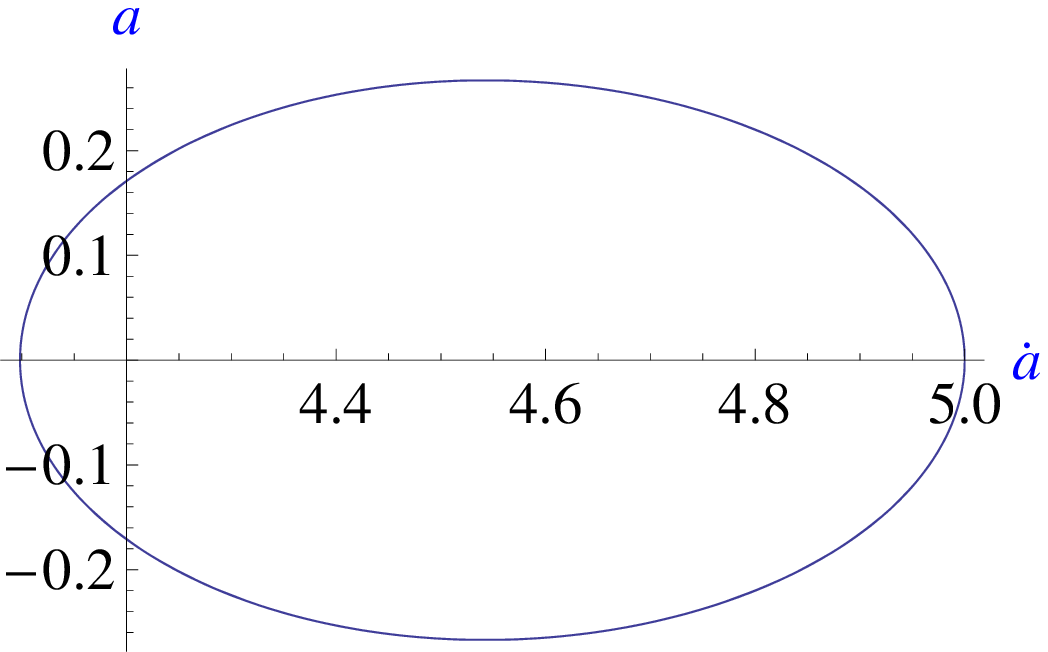}
 \caption{\label{Fig1} The evolutionary curve of the scale factor with time (left) and
the phase diagram  in space ($a$, $\dot{a}$) (right) for the case
$\Lambda<0$ in Planck unit  and with $w = -0.90$, $\Lambda = -0.6$,
$\alpha = -1$. }
\end{figure}

In this case, $\alpha<0$ and Point A is physically meaningless since
$a_{Es}^2<0$. Therefore, we only discuss the stability of Point B,
which  exists under the condition $-1<w<\frac{1}{3}$. From
Eq.~(\ref{eigen2}), it is easy to see  that $\vartheta^2<0$  since
$\alpha\Lambda>0$. This means that point $B$ is stable. In
Fig.~\ref{Fig1}, we plot the evolution of the scale factor with
time and the phase diagram in space ($a, \dot{a}$). This figure
shows that the Universe can stay at the stable state eternally and
may undergo a series of infinite, nonsingular oscillations about
this point. Thus, the initial big bang singularity can be avoided.
By numerical calculation, however, one finds that, when $w$ is
larger than $\frac{1}{3}$ or less than $-1$,  the Universe may
undergo an  accelerating expansion. It therefore appears that the
universe may enter an inflationary phase from this stable point if
the condition $-1<w<\frac{1}{3}$ is violated. However, from
Eq.~(\ref{S2}), we find that, once $w$ evolves through $-1$ or
$\frac{1}{3}$, the scale factor $a$  becomes $\infty$ or $0$.
Therefore,  in this case, the Universe is essentially stuck at the
stable static state unless the scale factor becomes singular.

\subsubsection{$\Lambda>0$}
A positive $\Lambda$ can be obtained by making an analytical
continuation of the parameters $\mu$ and $\omega^2$ by
$\mu\rightarrow i\mu$ and $\omega^2\rightarrow
i\omega^2$~\cite{Lu2009}. The analytical continuation  changes the
sign of $\alpha$ and makes it a positive constant ($\alpha>0$),
since it contains a $\mu^2$ factor. Now, critical point A can exist,
but it is a saddle point since $\vartheta^2>0$. The critical point B
is physically meaningless due to $\rho_B<0$ as we have pointed out.
Therefore, in this case, there is no stable Einstein static
universe.

A summary of the existence and stability of Points A and B is given in Table \ref{Tab1}.

\begin{table}[!h]
\tabcolsep 3pt \caption{\label{Tab1} Summary of the critical points and their stability in the case of the HL cosmology} \vspace*{-12pt}
\begin{center}
\begin{tabular}{|c|c|c|c|c|c|}
  \hline
      & \multicolumn{2}{|c|}{$\Lambda<0$}         & \multicolumn{2}{|c|}{$\Lambda>0$} \\ \cline{2-5}
      & Existence & Stability       & Existence & Stability \\ \hline
  Point A  & meaningless &         & $\forall w$ & unstable \\ \hline
  Point B  & $-1<w<\frac{1}{3}$ &stable    & meaningless &  \\
  \hline
\end{tabular}
       \end{center}
       \end{table}

\subsection{The SVW HL cosmology}
For the SVW HL cosmology,  we only consider the case of a positive
cosmological constant ($\Lambda>0$)\footnote{The negative
cosmological constant case can be treated similarly}. From
Eq.~(\ref{SVW3}), one can see that the critical points are
determined by the following cubic equation:
\begin{eqnarray}\label{SVWC}
(1+w)-\frac{1+3w}{\Lambda{a}^2}+\frac{6w-2}{\Lambda^2{a}^4}\bar{\beta}_1+\frac{12w-12}{\Lambda^3{a}^6}\bar{\beta}_2=0
\end{eqnarray}
When $\beta_2=0$, the above equation simplifies to a quadratic one,
which is similar with that in the HL theory, but not identical,
since  coefficients are different. Thus, now we separately  discuss
two cases, $\beta_2=0$ and $\beta_2\neq 0$.

\subsubsection{$\beta_2=0$}
In this case, Eq.~(\ref{SVWC}) reduces to
\begin{eqnarray}\label{SVWC2}
(1+w)-\frac{1+3w}{\Lambda{a}^2}+\frac{6w-2}{\Lambda^2{a}^4}\bar{\beta}_1=0.
\end{eqnarray}
Solving this equation, one can obtain  two critical points:
\begin{eqnarray}\label{PC}
Point
 \quad C \quad
 \frac{1}{\Lambda{a}^2_{Es}}=\frac{1}{4(3w-1)\bar{\beta}_1}[1+3w-\sqrt{(1+3w)^2-8\bar{\beta}_1(3w^2+2w-1)}]\;,
\end{eqnarray}
\begin{eqnarray}\label{PD}
Point
 \quad D  \quad \frac{1}{\Lambda{a}^2_{Es}}=\frac{1}{4(3w-1)\bar{\beta}_1}[1+3w+\sqrt{(1+3w)^2-8\bar{\beta}_1(3w^2+2w-1)}]\;.\end{eqnarray}
When $\beta_1=\frac{3}{8}$, critical points $C$ and $D$ reduce to $
\frac{1}{{a}^2_{Es}}=\frac{2}{3}\Lambda$ and
$\frac{1}{{a}^2_{Es}}=\frac{2(1+w)}{3w-1}\Lambda$, respectively,
which are the same as that in the HL cosmology [given in
Eqs.~(\ref{S1}, \ref{S2})] after a redefinition  of the
cosmological constant as $\frac{2}{3}\Lambda$.  It follows, from
Eqs.~(\ref{PC}, \ref{PD}), that when
\begin{eqnarray}\label{cc}\bar{\beta}_1=\frac{(1+3w)^2}{8(3w-1)(w+1)},\end{eqnarray} two critical points, C
and D, coincide, and thus, in this case,  there is only one critical
point,
\begin{eqnarray}\frac{1}{{a}^2_{Es}}=\frac{w+1}{4(3w+1)}\Lambda\;.\end{eqnarray}
The energy density $\rho$, at critical points C and D, is
\begin{eqnarray}\rho({a}_{Es})=\bigg(-\frac{1}{3}+\frac{1}{\Lambda{a}_{Es}^2}-2\bar{\beta}_1\frac{1}{\Lambda^2{a}_{Es}^4}\bigg)\Lambda\;,\end{eqnarray}
where $\frac{1}{{a}_{Es}^2}$ is given by (\ref{PC}) or (\ref{PD}).
In order to ensure these critical points to exist with  a physical
meaning, it is required that $a^2_{Es}>0$ and $\rho({a}_{Es})\geq0$.
This yields a region of existence in the ($w, \bar{\beta}_1$)
parameter space,

$\bullet$ For Point C:
    \begin{eqnarray}
    &&w\leq-\frac{1}{3}, \quad 0\leq\bar{\beta}_1\leq\frac{3}{8}\;,\nonumber\\
    &&-\frac{1}{3}\leq w\leq\frac{1}{3}, \quad
    \frac{(1+3w)^2}{8(3w^2+2w-1)}\leq\bar{\beta}_1\leq\frac{3}{8}\;,\nonumber\\
    &&w>\frac{1}{3}, \quad \bar{\beta}_1\leq\frac{3}{8}\;.
    \end{eqnarray}

\begin{figure}[htbp]
\includegraphics[width=5cm]{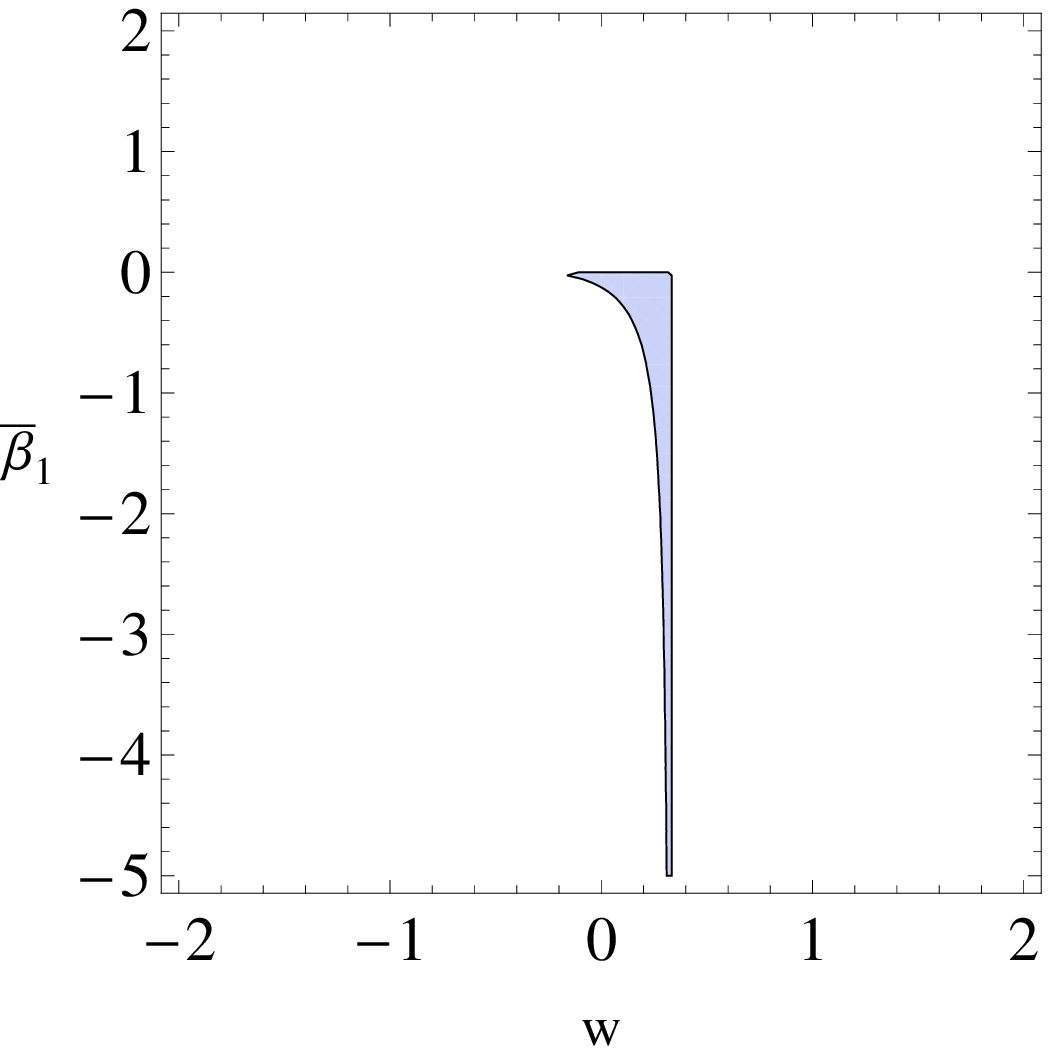}\quad\includegraphics[width=5cm]{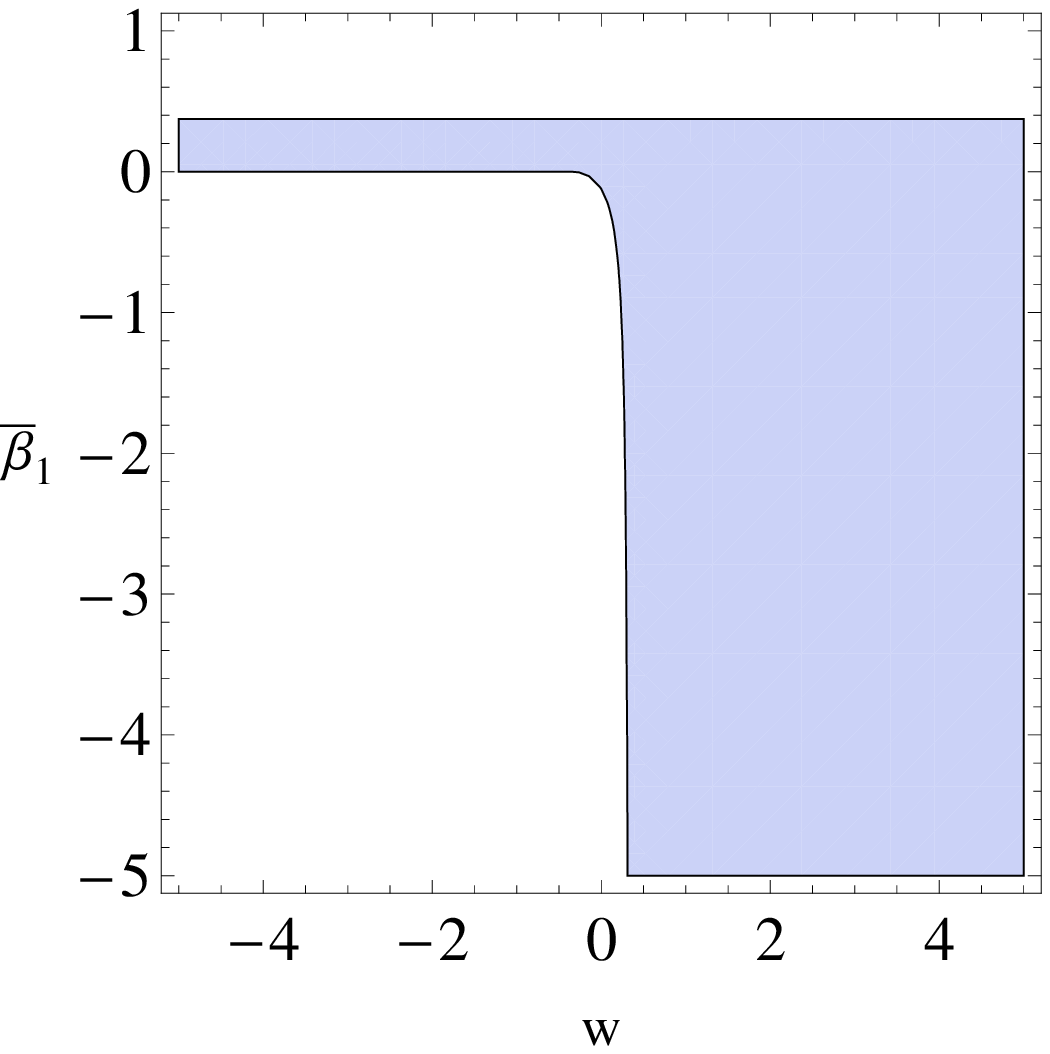}
 \caption{\label{Fig31} Regions of existence in the ($w, \bar{\beta}_1$) parameter space. The left panel shows the
 existence
region for Point $D$, while the right panel  for  Point $C$. }
\end{figure}

$\bullet$ For Point D:
    \begin{eqnarray}
    &&-\frac{1}{3}\leq w<\frac{1}{3}, \quad
    \frac{(1+3w)^2}{8(3w^2+2w-1)}<\bar{\beta}_1< 0\;.
    \end{eqnarray}
 In Fig.~\ref{Fig31}, we show the regions
of existence in the ($w, \bar{\beta}_1$) parameter space for both
critical points $C$ and $D$.

\begin{figure}[htbp]
\includegraphics[width=5cm]{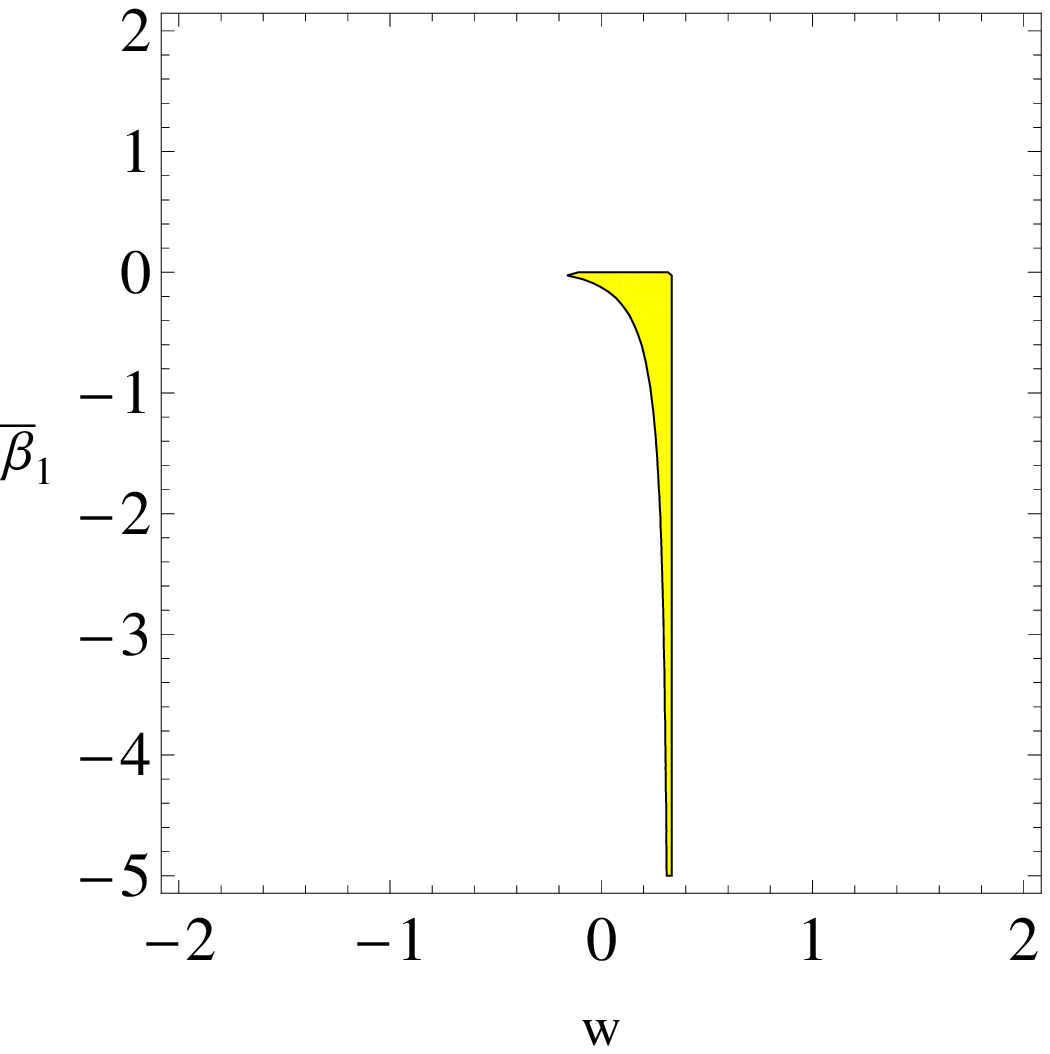}\quad\includegraphics[width=5cm]{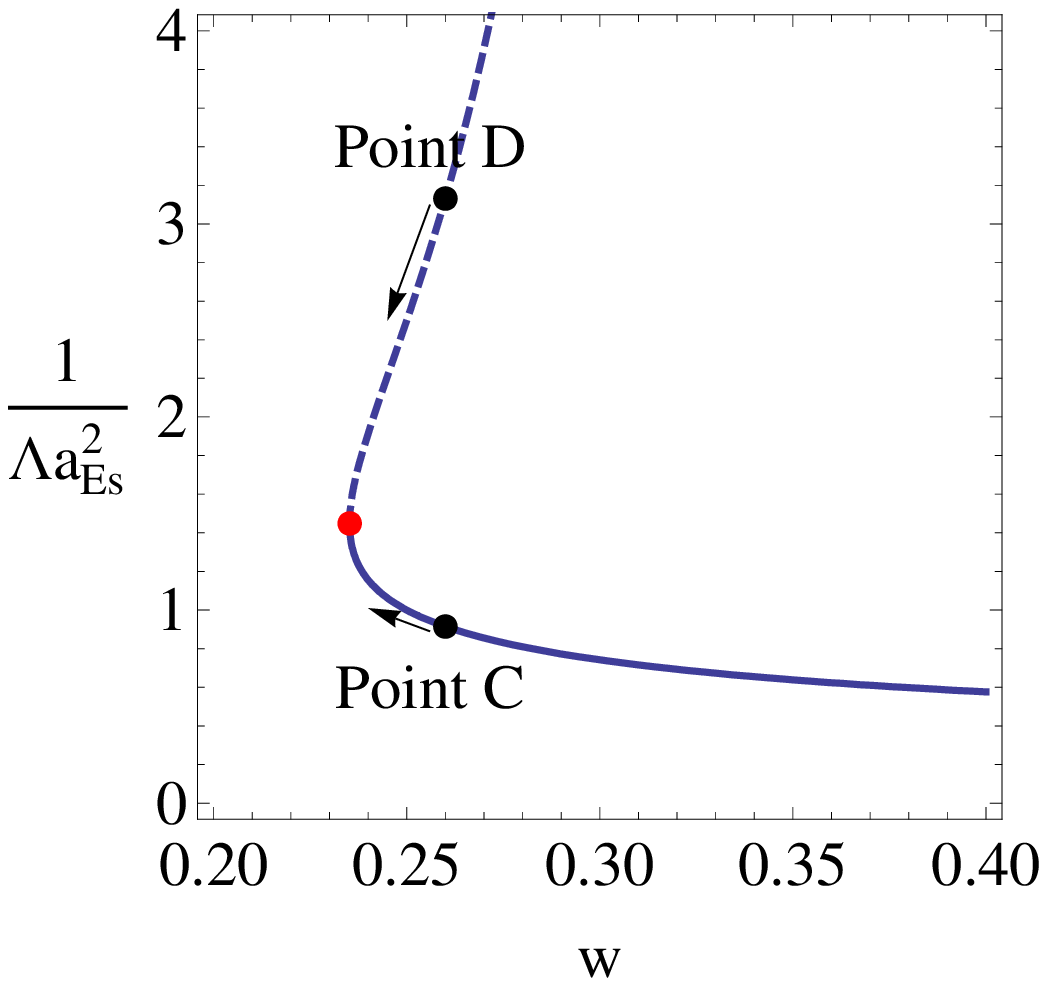}
 \caption{\label{Fig32}  Regions of stability  in the ($w, \bar{\beta}_1$) parameter space for Point $D$ (left panel),
and the evolution of Points C and D with the
 decreasing of $w$ (right panel).  In the right panel,  $\Lambda=0.6$ and $\bar{\beta}_1=-1$ }
\end{figure}

With the same method as that used in the  HL theory, we find the
eigenvalue of critical points $C$ and $D$, which can be expressed as
\begin{eqnarray}
\vartheta^2=(1+w)+\frac{1+3w}{\Lambda{a}_{Es}^2}-\frac{3(6w-2)\bar{\beta}_1}{\Lambda^2{a}_{Es}^4}\;.
\end{eqnarray}
There is no point in the existence region in the parameter ($w,
\bar{\beta}_1$) space  for Point $C$ which gives rise to a negative
$\vartheta^2$. Hence Point C is always unstable. For critical Point
$D$, the region of stability and existence is
\begin{eqnarray}\label{CPD}
    &&-\frac{1}{3}\leq w<\frac{1}{3}, \quad
    \frac{(1+3w)^2}{8(3w^2+2w-1)}<\bar{\beta}_1< 0\;,
    \end{eqnarray}
which means that, if Point $D$ exists, it is always stable. The left
panel of Fig.~\ref{Fig32} shows the region of parameters ($w,
\bar{\beta}_1$) corresponding to Point D. We summarize the existence and stability of points C and D in Table \ref{Tab2}.
\begin{figure}[htbp]
\includegraphics[width=6cm]{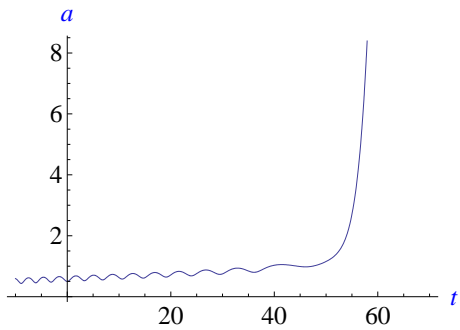}
 \caption{\label{Fig34}  The phase transition from a stable state to an inflation by assuming a slowly deceasing equation of state($w(t)=0.280-0.001t$) for the HL cosmology with the initial conditions $a(0)=0.5$ and $\dot{a}(0)=0$. The parameters are set as $\Lambda=0.6$ and $\bar{\beta}_1=-1$  }
\end{figure}

Thus, if the cosmic scale factor satisfies Eq.~(\ref{PD}) initially,
and $w$ and $\bar{\beta}_1$ lie in the region given in (\ref{CPD}),
the Universe can stay at a stable state past-eternally and undergo
an infinite oscillation. If $w$ evolves in such a way that  $w$ and
$\bar{\beta}_1$ satisfy Eq.~(\ref{cc}), then the stable critical
point $D$ coincides with the unstable one (Point C) and  becomes
unstable. As a result, the universe goes out of the stable state and
 enters  an inflationary phase naturally. A particular case which realizes a phase transition  from a stable state to an
 inflation era
 is shown in Fig.~\ref{Fig34}. So the big rip singularity
 may be avoided successfully in this case.

\begin{table}[!h]
\tabcolsep 3pt \caption{\label{Tab2} Summary of the critical points and their stability in the SVW HL cosmology with $\beta_2=0$} \vspace*{-12pt}
\begin{center}
\begin{tabular}{|c|c|c|c|c|c|}
  \hline
           & Existence               & Stability        \\ \hline
  Point C  & $w\leq-\frac{1}{3},\;0\leq\bar{\beta}_1\leq\frac{3}{8}$ &   unstable       \\
           & $-\frac{1}{3}\leq w\leq\frac{1}{3}, \; \frac{(1+3w)^2}{8(3w^2+2w-1)}\leq\bar{\beta}_1\leq\frac{3}{8}$ &    \\
           & $w>\frac{1}{3}, \quad \bar{\beta}_1\leq\frac{3}{8}$&                    \\ \hline
  Point D  & $-\frac{1}{3}\leq w<\frac{1}{3}, \;  \frac{(1+3w)^2}{8(3w^2+2w-1)}<\bar{\beta}_1< 0$ &  stable   \\
  \hline
\end{tabular}
       \end{center}
       \end{table}

\subsubsection{$\beta_2\neq 0$}
Now, the Einstein static points satisfy  Eq.~(\ref{SVWC}), which is
a cubic equation of ${a}^2_{Es}$. The solution of Eq.~(\ref{SVWC})
is determined by the following expression
\begin{eqnarray}
\Delta=B^2-4AC\;,
\end{eqnarray}
where $A=b^2-3ac$, $B=bc-9ad$, and $C=c^2-3bd$ with
$a\equiv12(w-1)\bar{\beta}_2$, $b\equiv2(3w-1)\bar{\beta}_1$,
$c\equiv-(1+3w)$ and $d\equiv (1+w)$.

$\bullet$  $\Delta>0$: there is only one real solution, which
corresponds to only one critical point:
\begin{eqnarray}\label{PE}
Point\quad E:\quad
\frac{1}{\Lambda{a}^2_{Es}}=-\frac{1}{3a}[b+Y_1^{1/3}+Y_2^{1/3}]\;,
\end{eqnarray}
where $Y_{1,2}=Ab+\frac{3a}{2}(-B\pm \sqrt{\Delta})$

$\bullet$ $\Delta<0$: there are three different real solutions. Thus, in this case, there are
three critical points:
\begin{eqnarray}\label{PF}
Point\quad F:\quad
\frac{1}{\Lambda{a}^2_{Es}}=-\frac{1}{3a}[b+2\sqrt{A}\cos(\theta/3)]\;,
\end{eqnarray}
\begin{eqnarray}\label{PG}
Point\quad G:\quad
\frac{1}{\Lambda{a}^2_{Es}}=-\frac{1}{3a}\big(b-\sqrt{A}[\cos(\theta/3)+\sqrt{3}\sin(\theta/3)]\big)\;,
\end{eqnarray}
\begin{eqnarray}\label{PH}
Point\quad H:\quad
\frac{1}{\Lambda{a}^2_{Es}}=-\frac{1}{3a}\big(b-\sqrt{A}[\cos(\theta/3)-\sqrt{3}\sin(\theta/3)]\big)\;,
\end{eqnarray}
where $\theta=\arccos(T)$ and $T=\frac{1}{2A^{3/2}}(2Ab-3aB)$.

$\bullet$ $\Delta=0$:  Points G  and H coincide since $\theta=0$ and thus there are
two critical points (Points F and G with $\theta=0$).

At these critical points, the corresponding energy density $\rho$
has the form
\begin{eqnarray}
\rho({a}_{Es})=-\frac{1}{3}+\frac{1}{\Lambda{a}^2_{Es}}-2\bar{\beta}_1\frac{1}{\Lambda^2{a}^4_{Es}}-4\bar{\beta}_2\frac{1}{\Lambda^3{a}^6_{Es}}\;,
\end{eqnarray}
with $\frac{1}{{a}_{Es}^2}$  given in Eq.~(\ref{PE})-(\ref{PG}), or (\ref{PH}). The conditions for these points to be
physically meaningful are that $\rho({a}_{Es})\geq 0$ and
${a}_{Es}^2>0$. Since it is not an easy task to obtain  analytic
solutions to the existence  conditions,  we resort to numerical
calculations and find that in order to satisfy the existence
conditions, for Points F, G and H, $\bar{\beta}_1$ is required to be
 less than about $\frac{3}{4}$, while  for Point E, there is no
constraint on $\bar{\beta}_1$.

\begin{figure}[htbp]
\includegraphics[width=5cm]{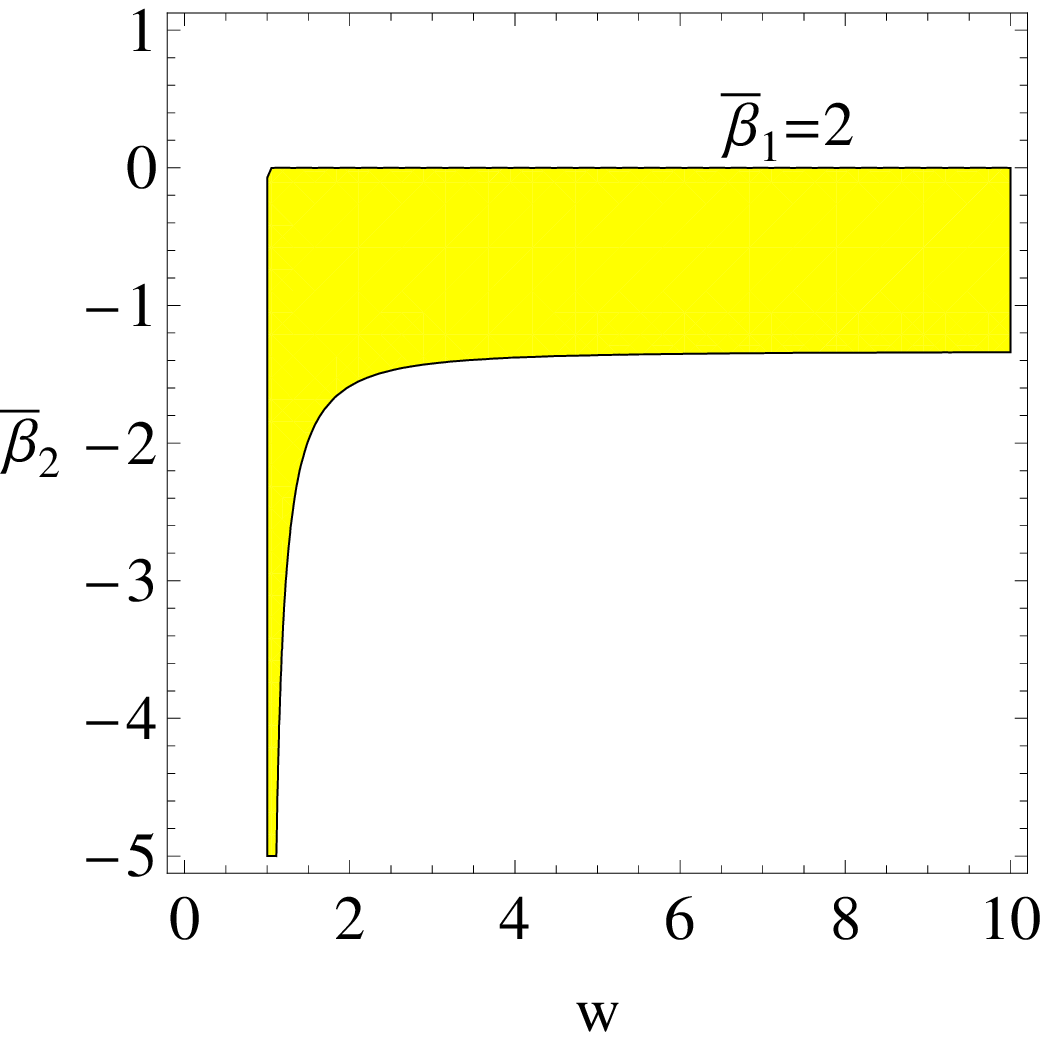}\quad\includegraphics[width=5cm]{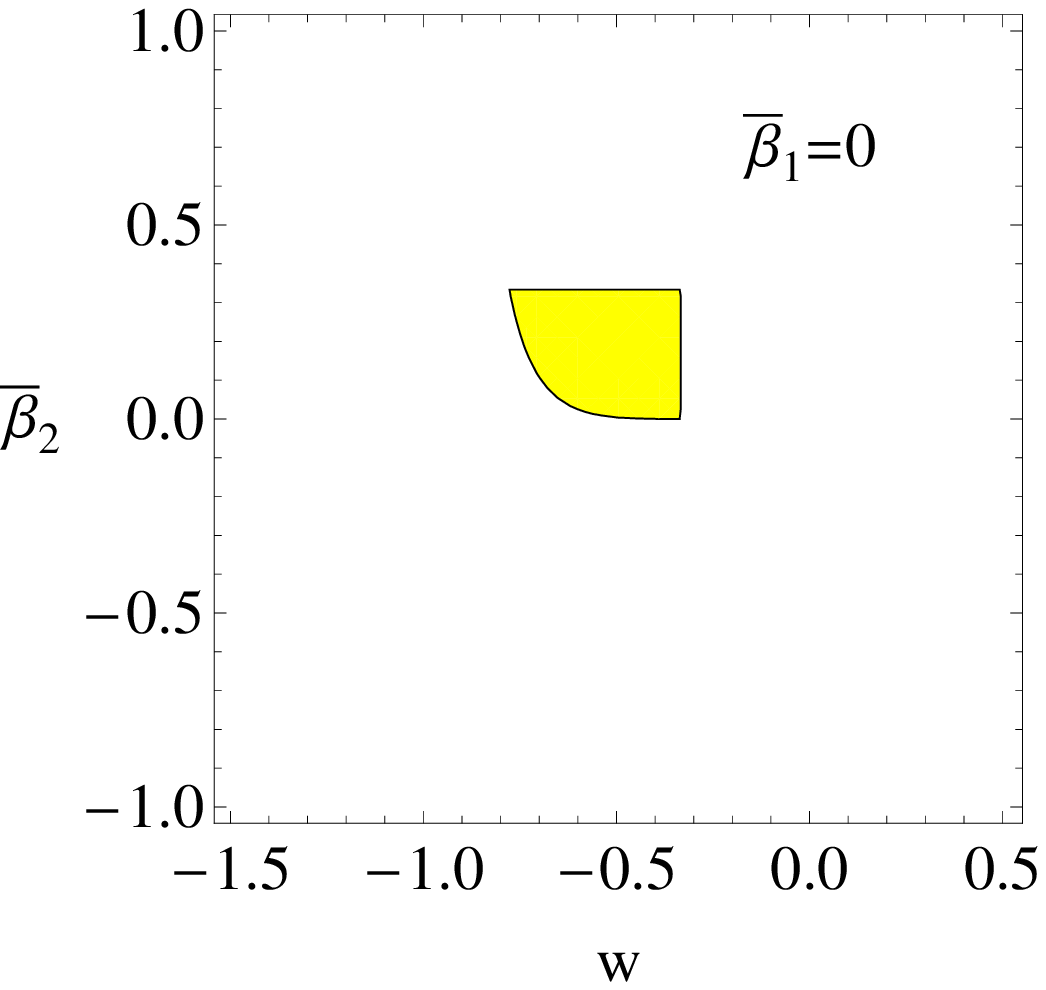}\quad\includegraphics[width=5cm]{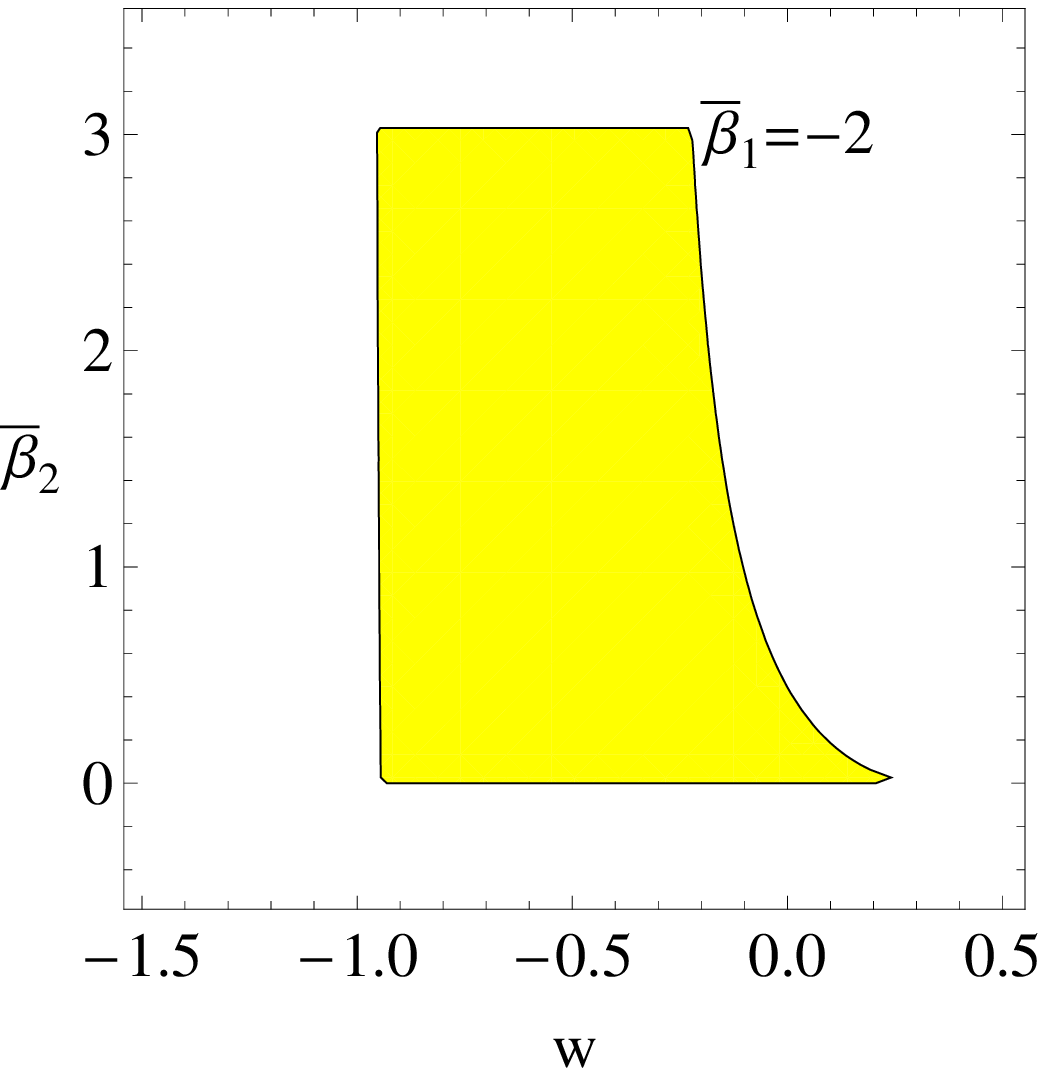}
 \caption{\label{Fig41} Regions of existence for Point E in the ($w, \bar{\beta}_2$) parameter space.
The left, middle and right panels show the case with
$\bar{\beta}_1=2$, $0$ and $-2$, respectively.}
\end{figure}

\begin{figure}[htbp]
\includegraphics[width=5cm]{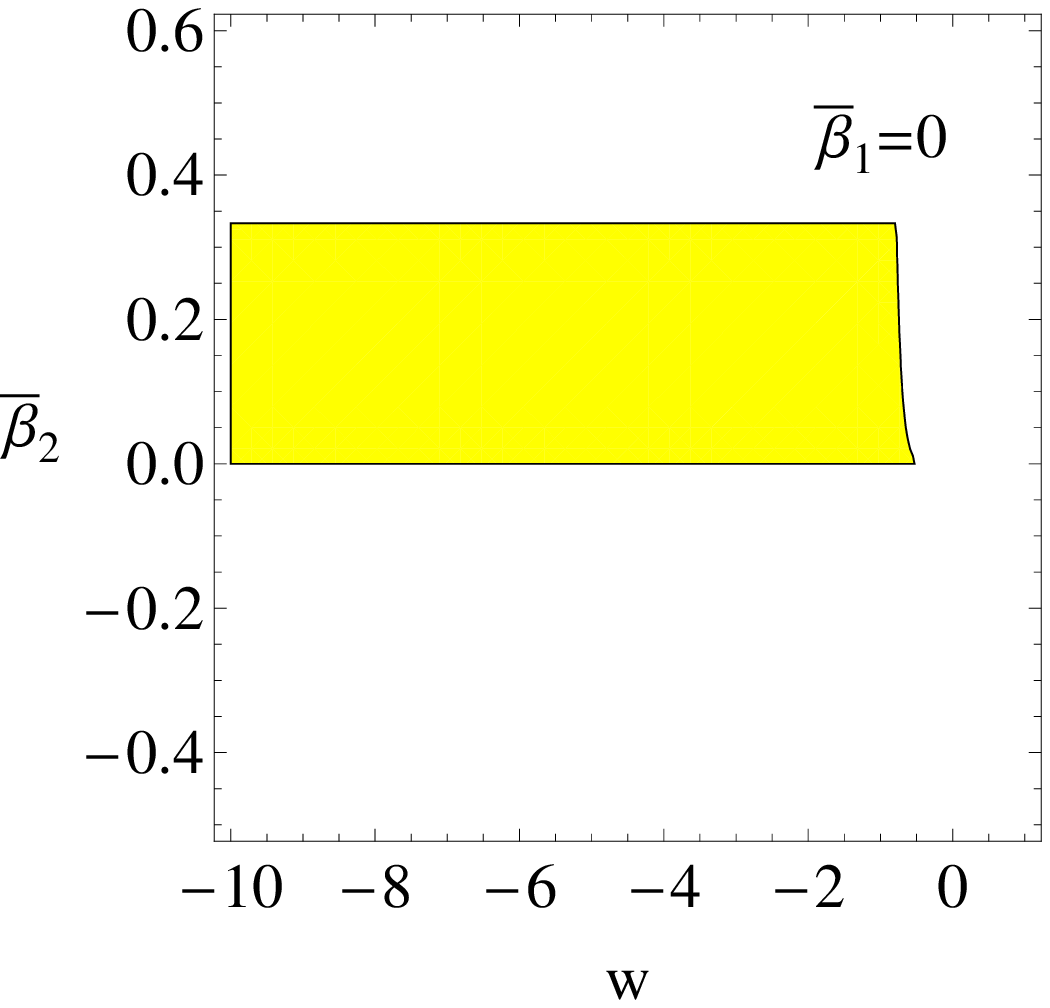}\quad\includegraphics[width=5cm]{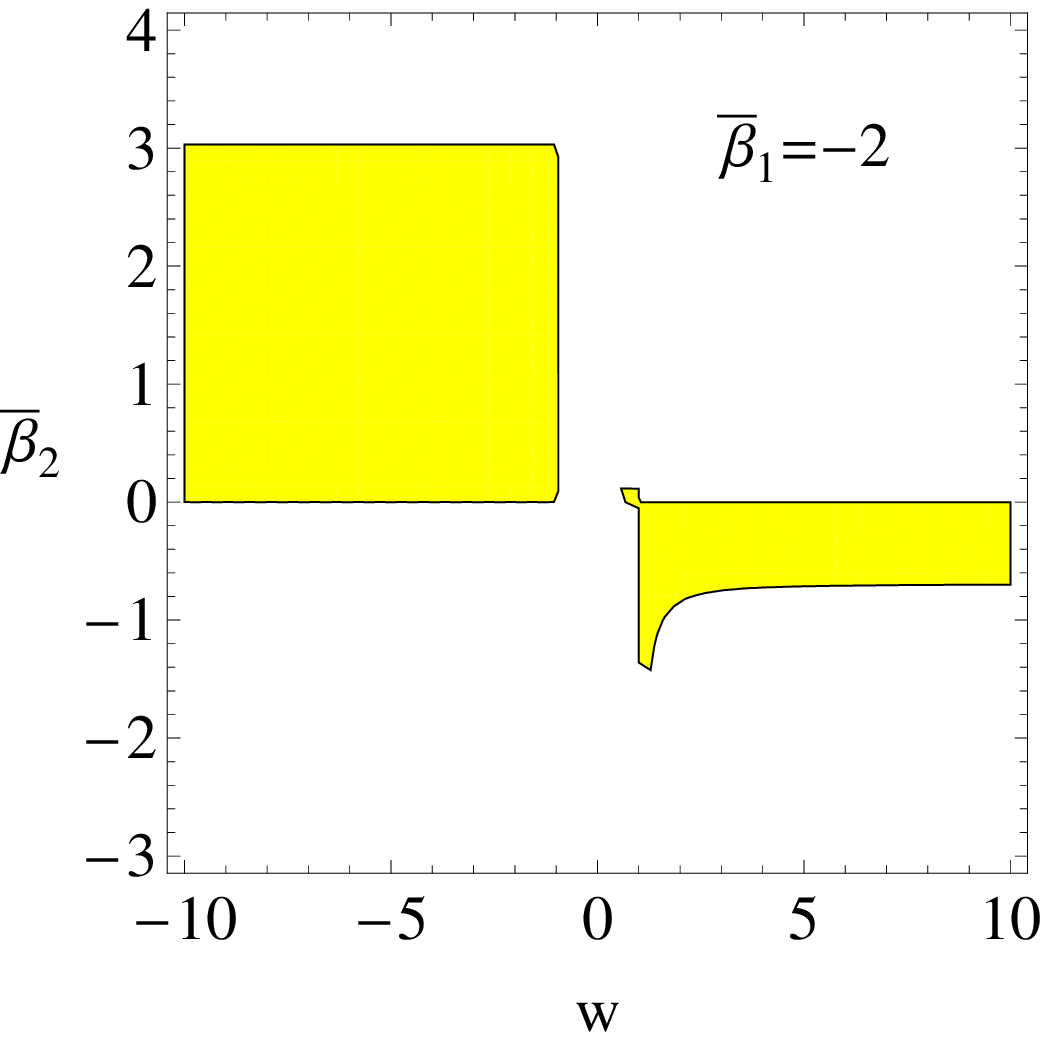}
 \caption{\label{Fig51} Regions of existence for Point F in the ($w, \bar{\beta}_2$) parameter space.
The left and right panels show the case with
$\bar{\beta}_1=0$ and $-2$, respectively.}
\end{figure}

\begin{figure}[htbp]
\includegraphics[width=5cm]{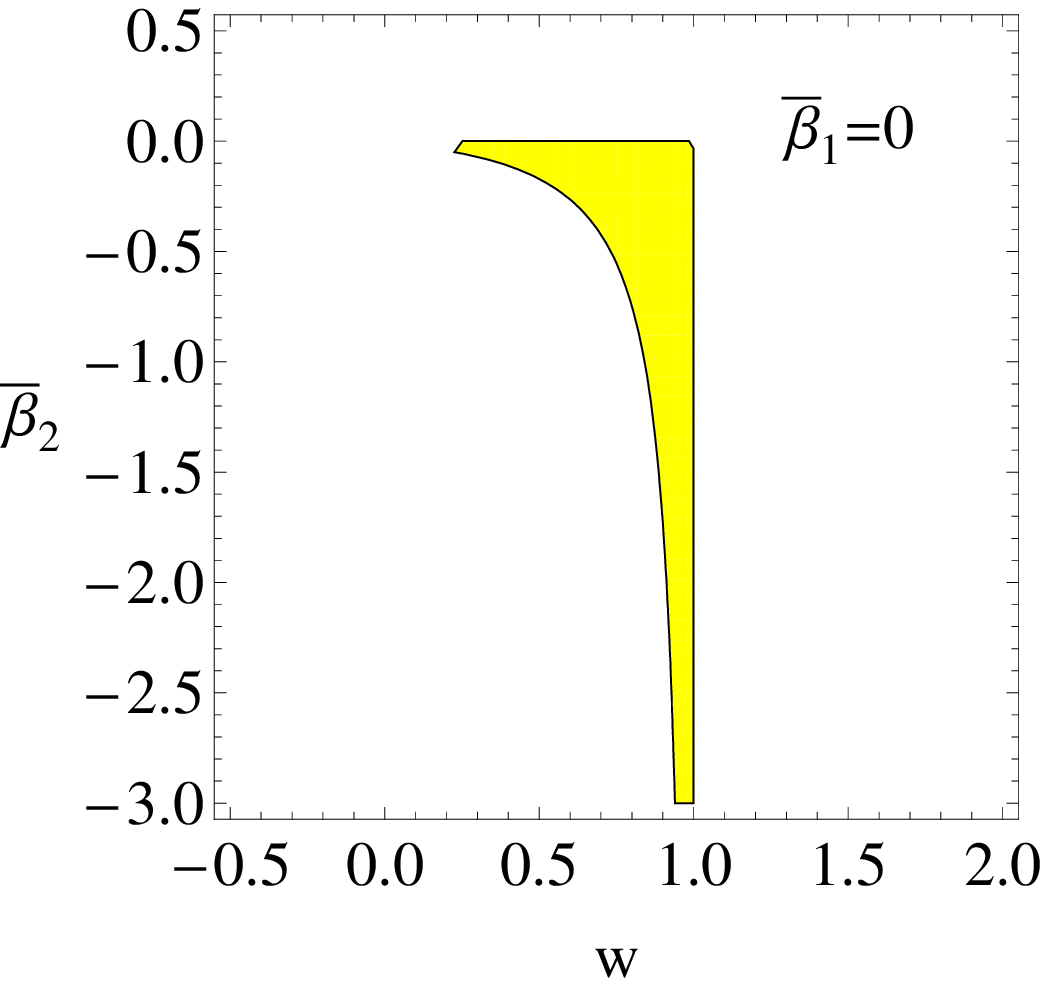}\quad\includegraphics[width=5cm]{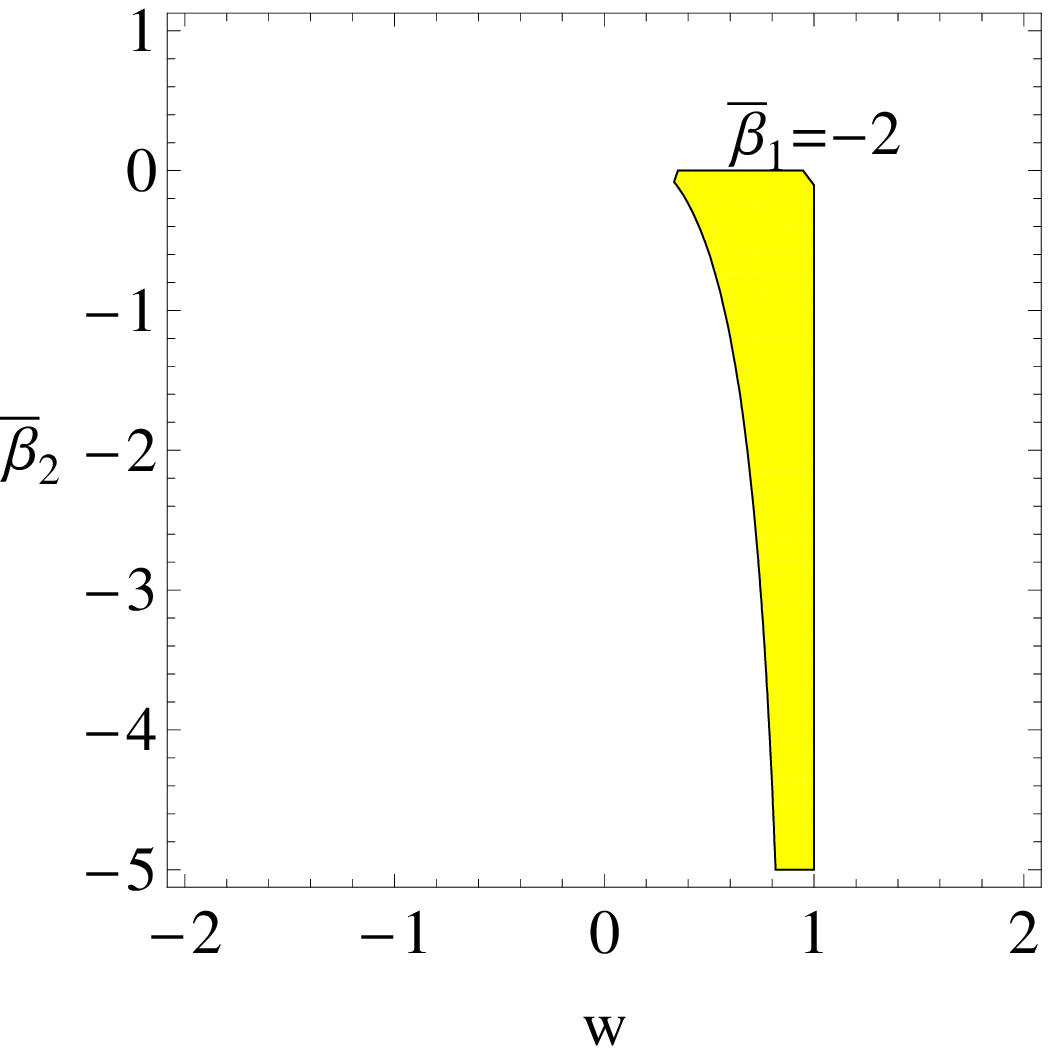}
 \caption{\label{Fig61} Regions of existence for Point G in the ($w, \bar{\beta}_2$) parameter space.
The left and right panels show the case with
$\bar{\beta}_1=0$ and $-2$, respectively.}
\end{figure}

\begin{figure}[htbp]
\includegraphics[width=5cm]{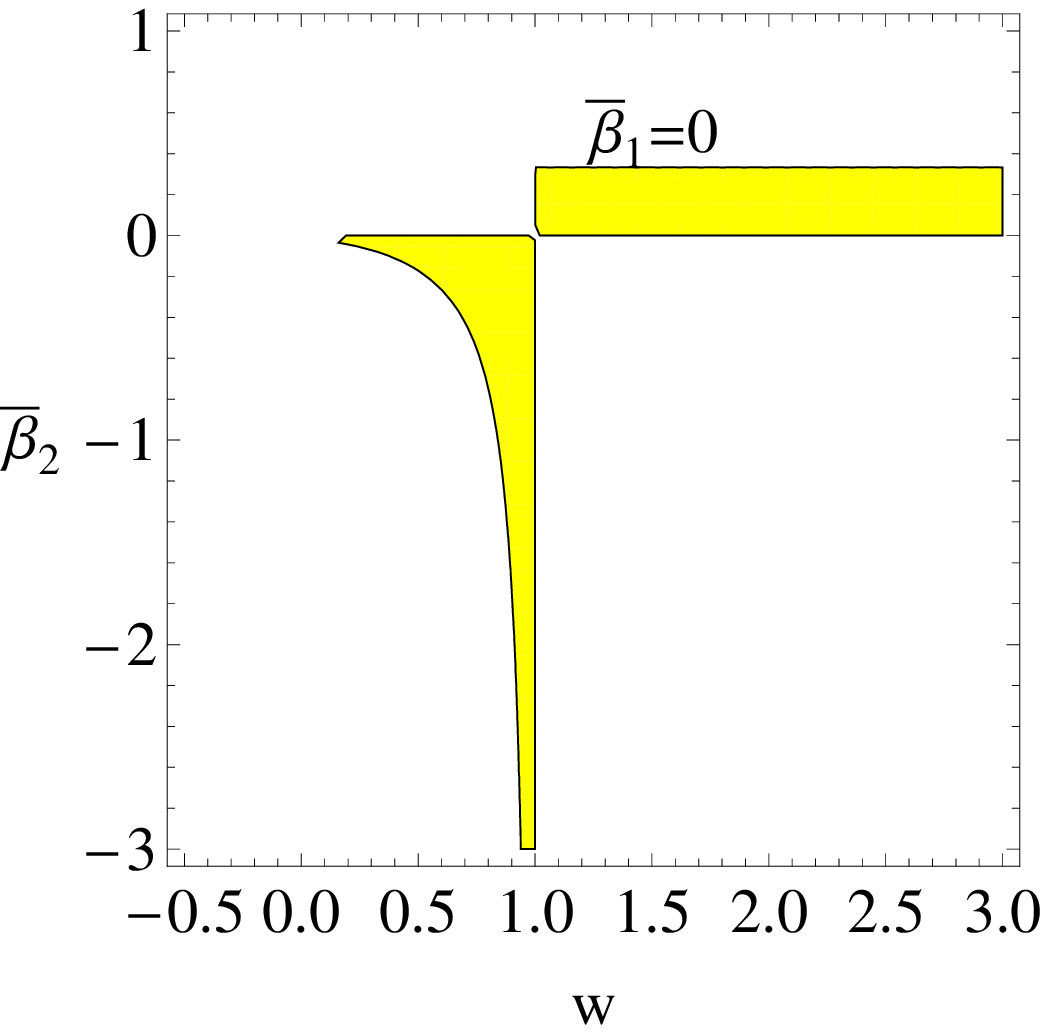}\quad\includegraphics[width=5cm]{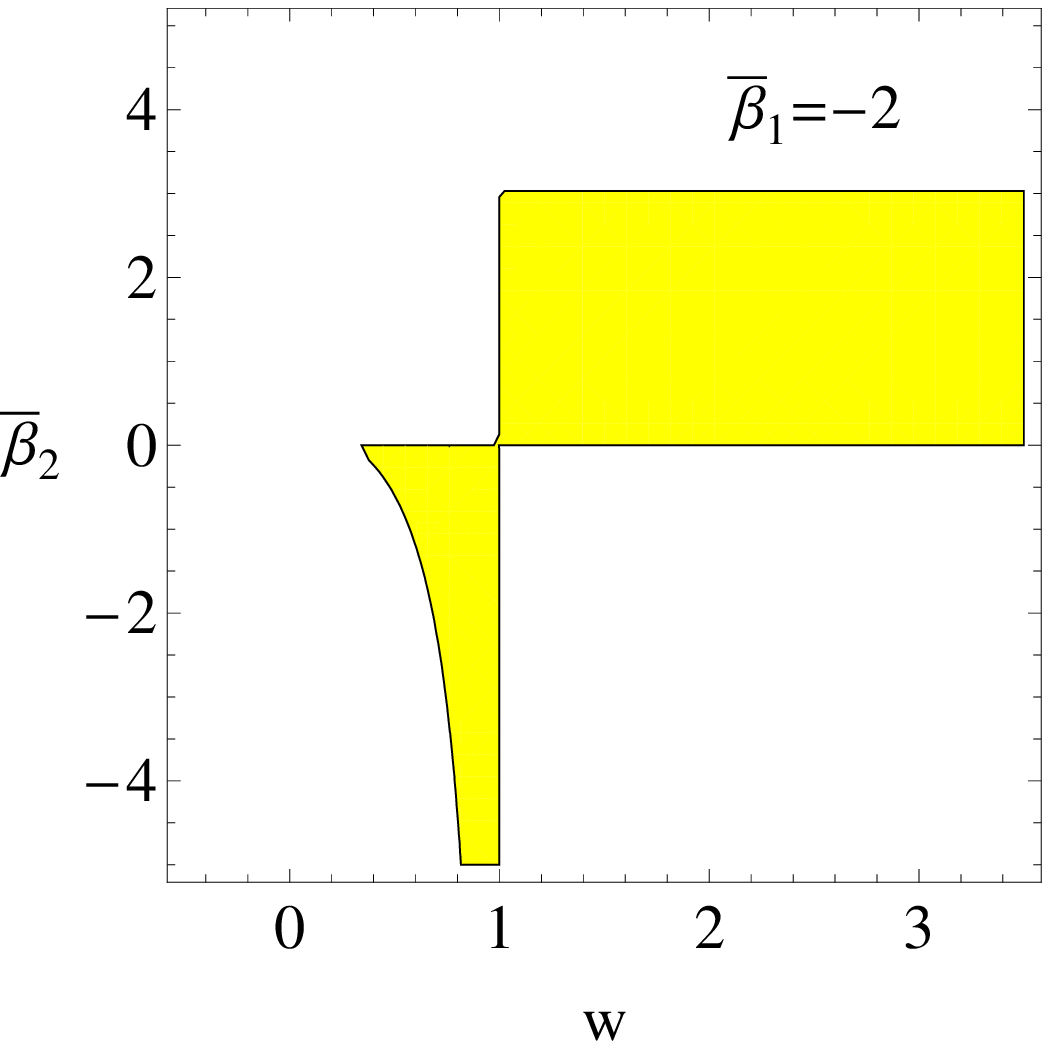}
 \caption{\label{Fig71} Regions of existence for Point H in the ($w, \bar{\beta}_2$) parameter space.
The left and right panels show the case with $\bar{\beta}_1=0$ and
$-2$, respectively.}
\end{figure}

In order to show the regions of existence for Points E, F, G and H
in the ($w$, $\bar{\beta}_2$) parameter space in detail, we chose
$-2$, $0$ and $2$ as three typical values  for $\bar{\beta}_1$.  The
results are shown graphically in Figs.~\ref{Fig41}-\ref{Fig71}.  Figure~\ref{Fig41} shows the regions of
existence for Point E with $\bar{\beta}_1=2, 0$ and $-2$,
Fig.~\ref{Fig51},  the region of existence for Point F with
$\bar{\beta}_1=0$ and
 $-2$, and Figs.~\ref{Fig61} and \ref{Fig71} the regions of existence
for Points G and H with $\bar{\beta}_1=0$ and
 $-2$.

\begin{figure}[htbp]
\includegraphics[width=5cm]{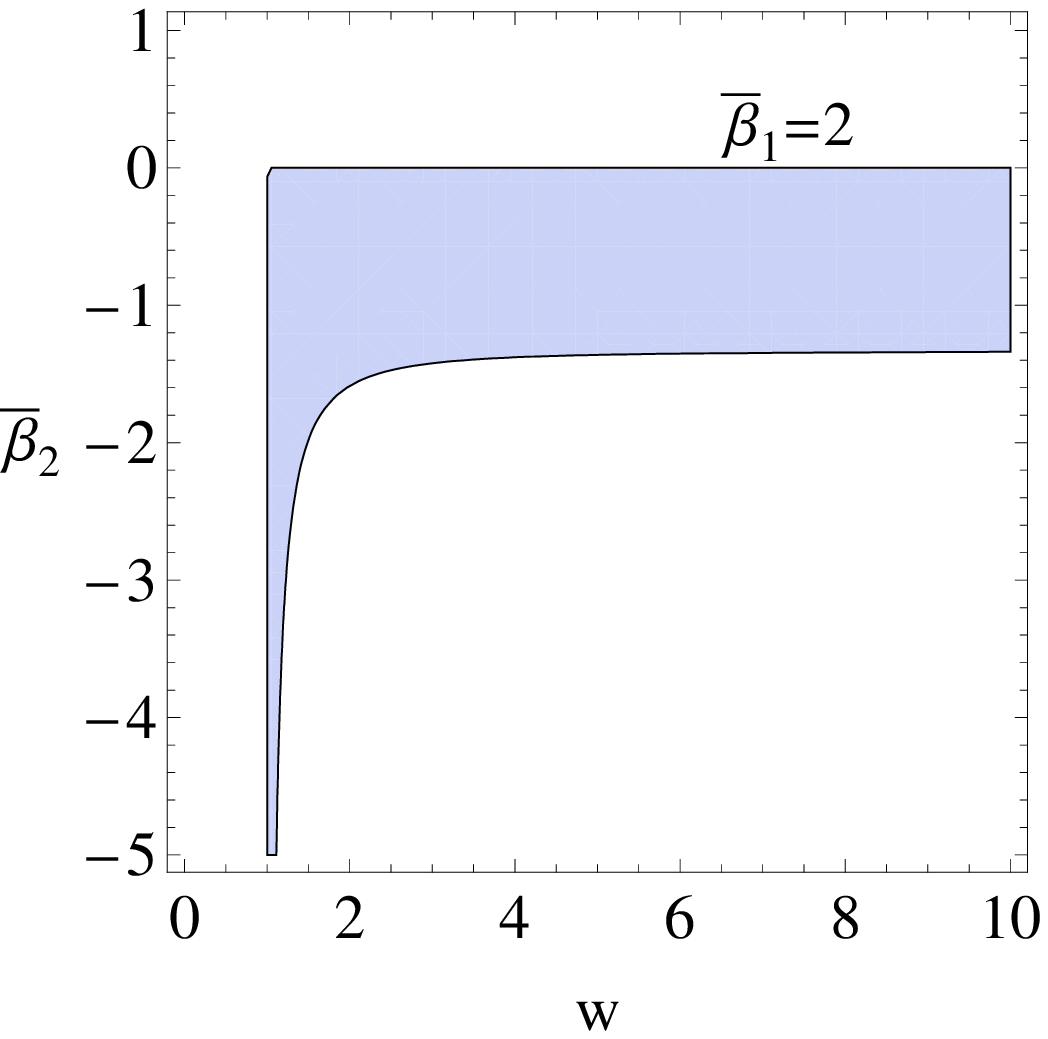}\quad\includegraphics[width=5.2cm]{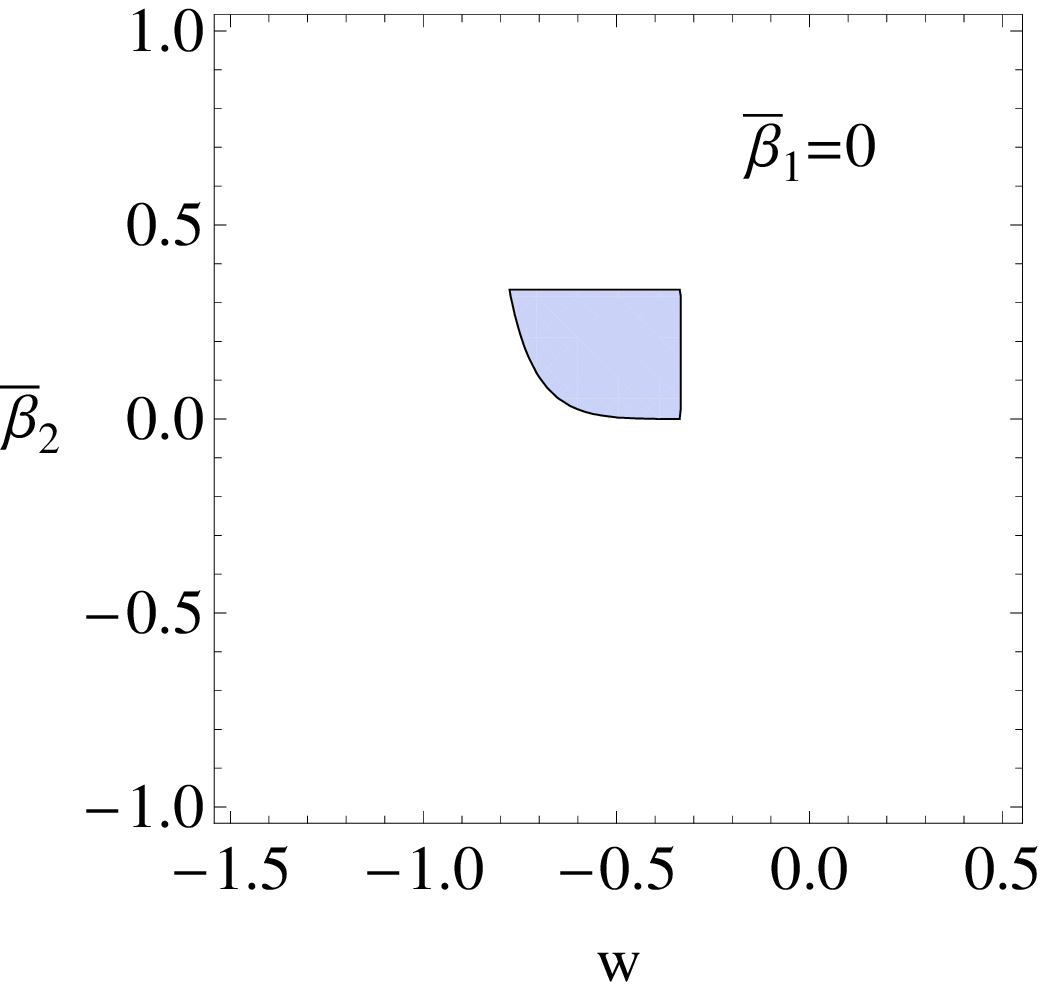}\quad\includegraphics[width=4.8cm]{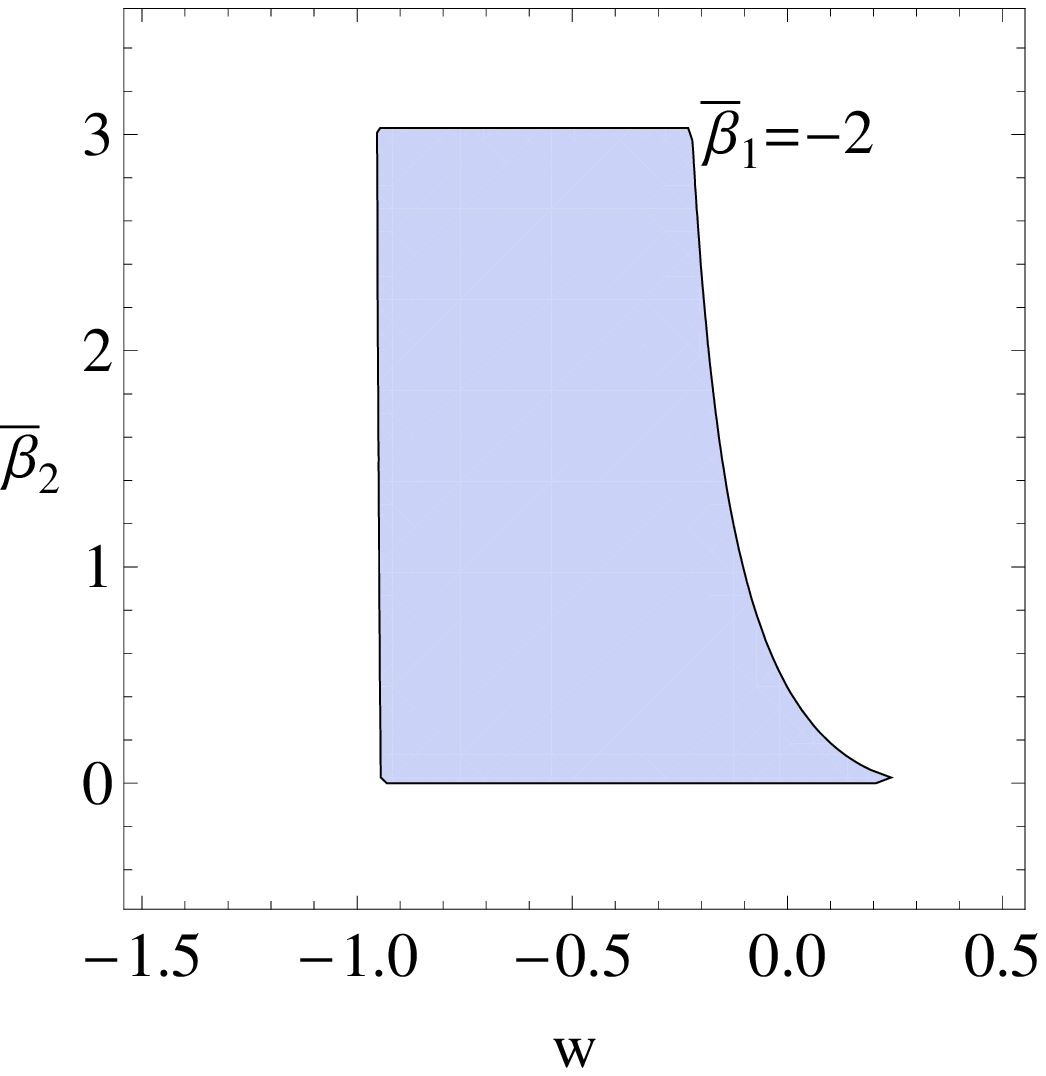}
 \caption{\label{Fig81} Regions of stability for Point E in the ($w, \bar{\beta}_2$) parameter space.
The left, middle and right panels show the case with
$\bar{\beta}_1=2$, $0$ and $-2$, respectively.}
\end{figure}

\begin{figure}[htbp]
\includegraphics[width=5cm]{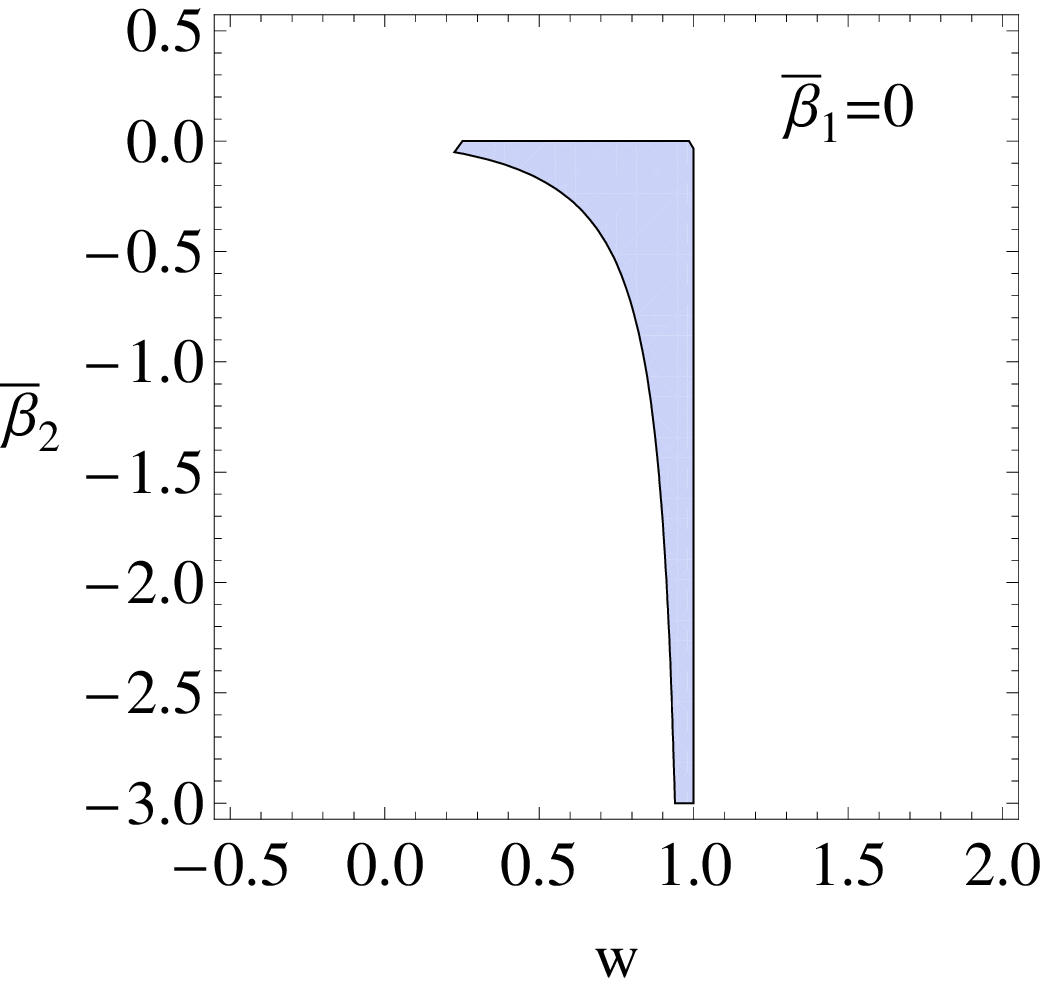}\quad\includegraphics[width=5cm]{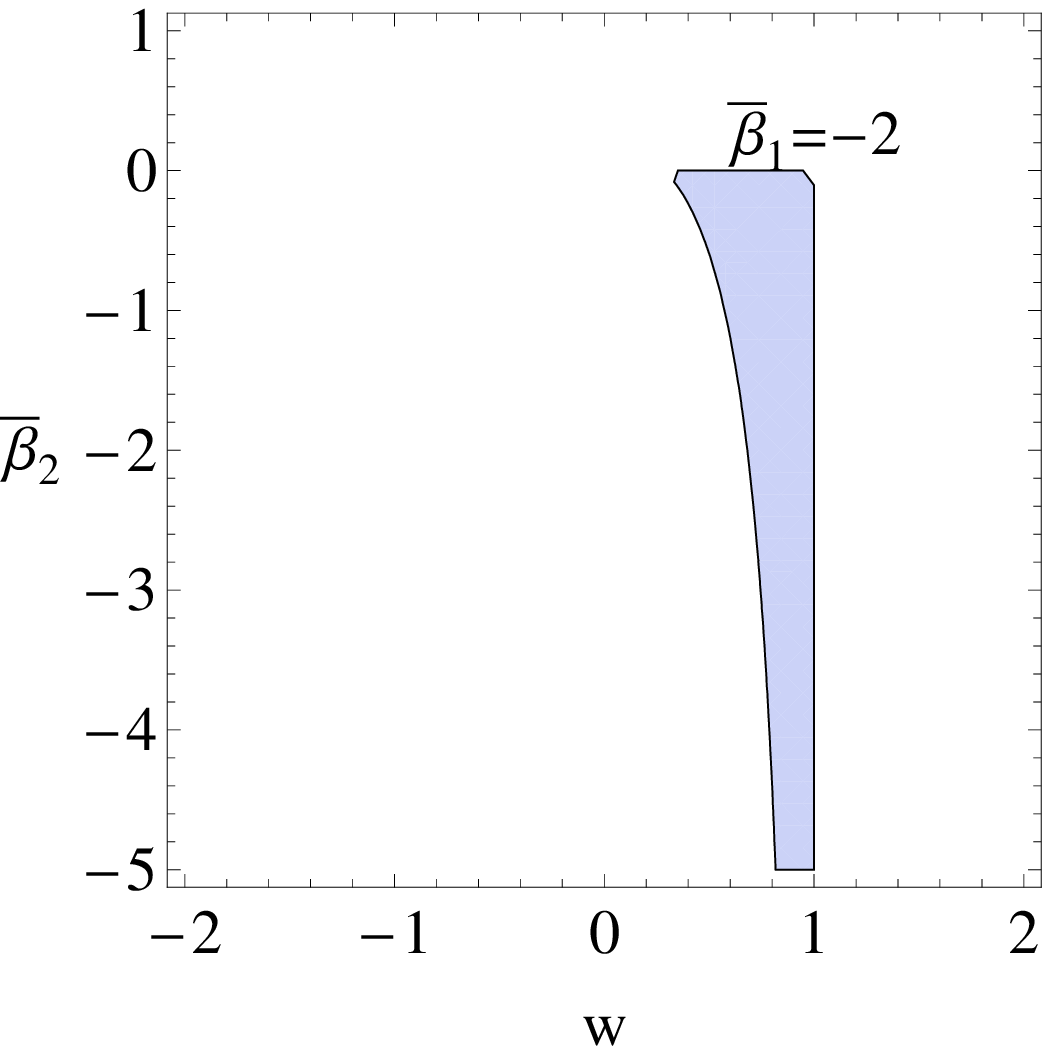}
 \caption{\label{Fig91} Regions of stability for Point G in the ($w, \bar{\beta}_2$) parameter space.
The left and right panels show the case with
$\bar{\beta}_1=0$ and $-2$, respectively.}
\end{figure}

In order to discuss the stability of these critical points, we need
to calculate the eigenvalues, which can be expressed as
\begin{eqnarray}
\vartheta^2=-\frac{1}{3}+\frac{1}{\Lambda{a}^2_{Es}}-2\bar{\beta}_1\frac{1}{\Lambda^2{a}^4_{Es}}-4\bar{\beta}_2\frac{1}{\Lambda^3{a}^6_{Es}}\;,
\end{eqnarray}
at each point. We find that the critical points F and H are always
unstable, while critical points, E and G,  are always stable as
long as they exist. A summary of these critical points and their stability is shown in Table \ref{Tab3}.
   Therefore, if the initial condition is such
that the cosmic scale factor satisfies Eq.~(\ref{PE}) or (\ref{PG}),
the big bang singularity can be avoided. To illustrate the stability
of Points E and G visually,   we plot the regions of stability in
$(w, \bar{\beta}_2)$ parameter space in Figs.~\ref{Fig81} and
\ref{Fig91}, respectively.

\begin{figure}[htbp]
\includegraphics[width=5cm]{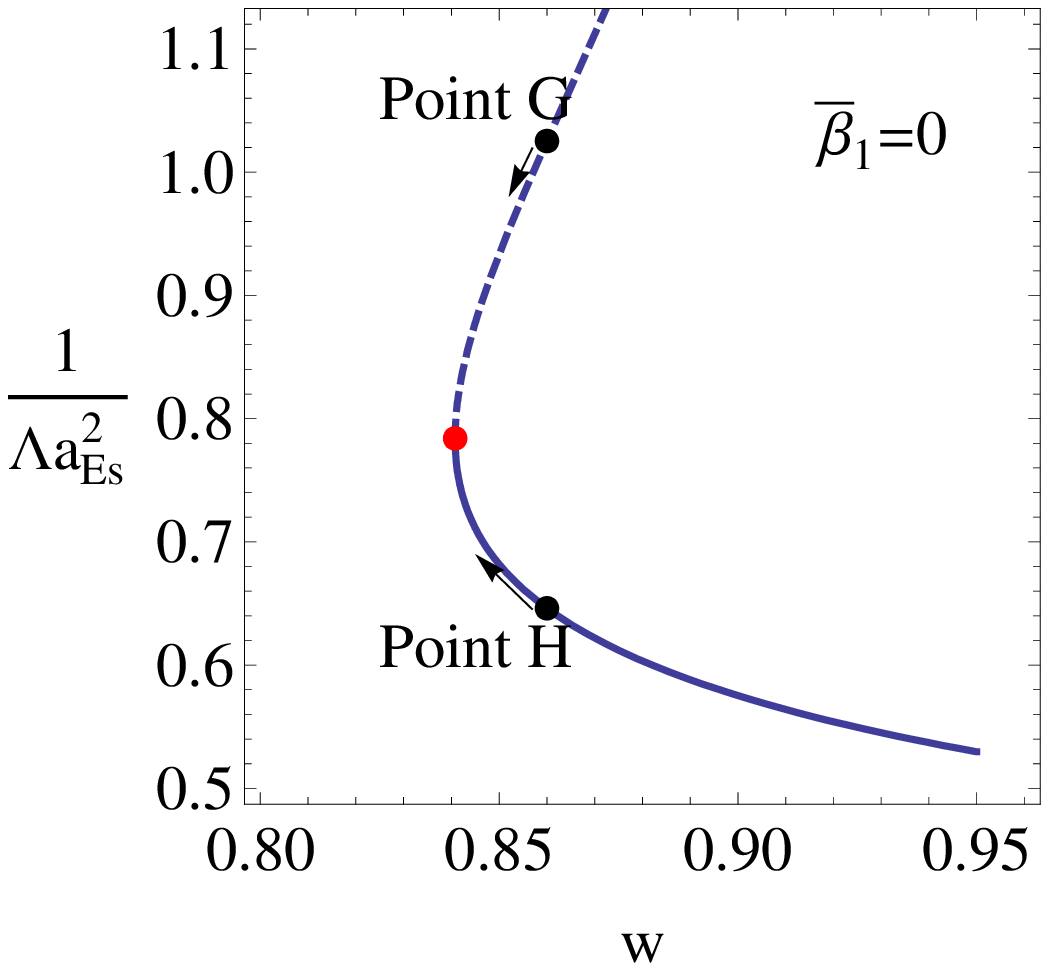}\quad\includegraphics[width=5.1cm]{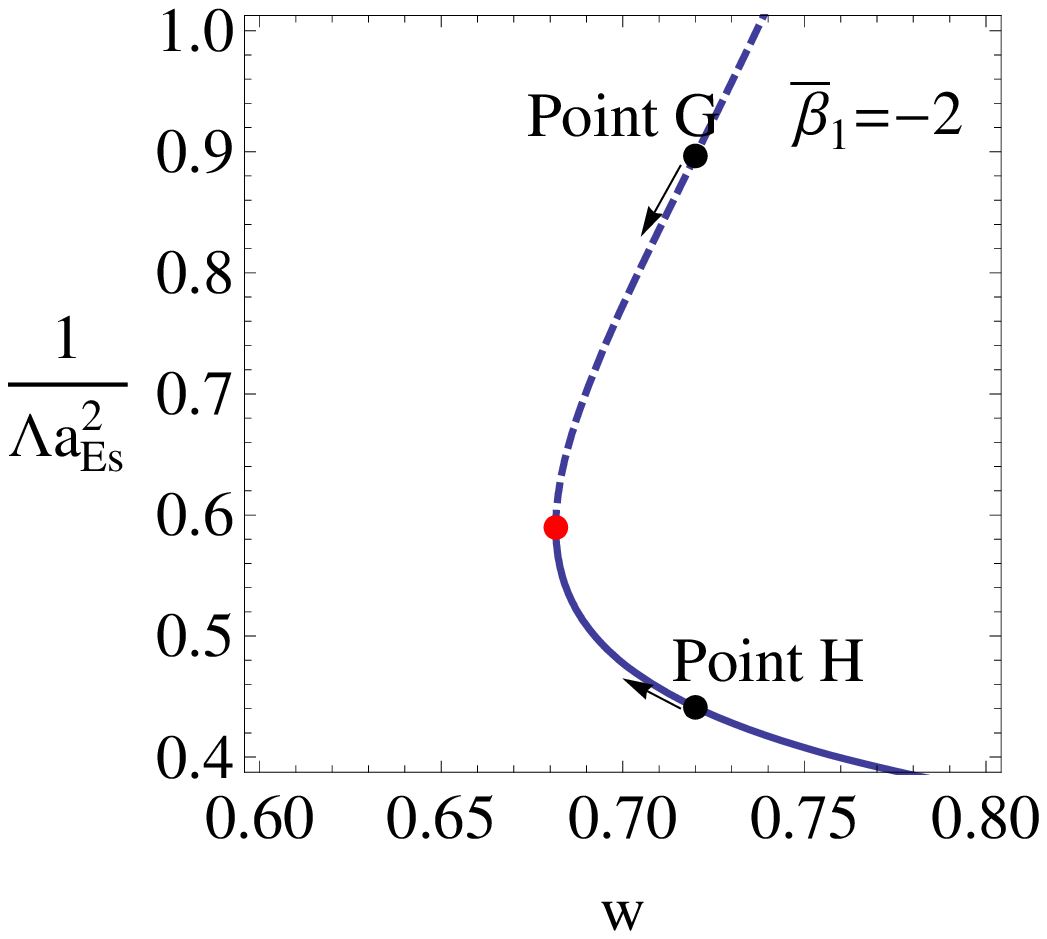}
 \caption{\label{Fig11} The evolutions of the stable Point G and unstable one
 H with the
 decreasing of $w$. }
\end{figure}

Let us note that  Points G and H move closer and closer as $\Delta$
changes from $\Delta<0$ to $0$, and coincide once $\Delta=0$, which
means that the stable Point G becomes an unstable one. Hence if the
cosmic scale factor satisfies the condition given in Eq.~(\ref{PG})
initially, the universe can evolve slowly from a stable region to an
unstable one with the decrease of $w$, as shown in the
Fig.~\ref{Fig11}. Because of the fact that this unstable state will
lead the Universe to enter an inflationary phase, therefore, in this
case, the big bang singularity can be avoided and a subsequent
inflation can appear naturally. Note, however, that if the cosmic
scale factor satisfies the condition given in Eq.~ (\ref{PE})
initially, the universe can stay at the Einstein static state
eternally and thus avoid the big bang singularity, but it cannot
evolve to an inflationary phase with the evolution of $w$.

\begin{table}[!h]
\tabcolsep 3pt \caption{\label{Tab3} Summary of the critical points and their stability in the SVW HL cosmology with $\beta_2\neq 0$} \vspace*{-12pt}
\begin{center}
\begin{tabular}{|c|c|c|c|c|c|c|c|}
  \hline
               & Critical point & Stability        \\ \hline
 $\triangle>0$ & Point E        & stable if it exists      \\ \hline
 $\triangle<0$ & Point F        &  unstable                 \\ \cline{2-3}
               & Point G        & stable if it exists       \\ \cline{2-3}
               & Point H        &  unstable                 \\ \hline
 $\triangle=0$ &  Point G or H with $\theta=0$ &  unstable                  \\ \cline{2-3}
              & Point F with $\theta=0$    &     unstable    \\   \hline
\end{tabular}
       \end{center}
       \end{table}

\section{Conclusion}
The Ho\v{r}ava-Lifshitz gravity is a power-counting renormalizable
theory,  which has an anisotropic scaling between space and time in
the UV limit,  and thus breaks the Lorentz invariance. Applying this
theory to cosmology,  one finds that  the Friedmann equation for a
nonflat universe  is modified by  a $\frac{1}{a^4}$ term. The SVW
HL theory is a generalization of the original HL gravity by keeping
the projectability condition but abandoning the detailed-balance
one. This generalization introduces an extra $\frac{1}{a^6}$
correction term to the Friedmann equation and  modifies the
coefficient for the $\frac{1}{a^4}$ term as  compared with the HL
theory. In the present paper, we study the influence of these
correction terms on the Einstein static state. In the case of HL
cosmology, if the cosmological constant $\Lambda$ is negative, we
find that there exists a stable Einstein static state. The Universe
can stay at this stable state eternally and thus the big bang
singularity can be avoided. However, in this case,  the universe can
not exit to an inflationary era. So the big bang singularity problem
cannot be solved successfully. By making an analytical continuation
of the model parameters, a positive $\Lambda$ can be
obtained~\cite{Lu2009}. But, in this case, there is no stable
Einstein static state.

For the SVW HL cosmology, when $\beta_2=0$,  we find that there
exists  a stable critical point and an unstable one. If the cosmic
scale satisfies the condition given in Eq.(\ref{PD}) initially, the
universe can stay at the Einstein static state past eternally. With
the decrease of $w$, the stable point and the saddle one move closer
and closer. Once $w$ reaches a critical value, this stable critical
point coincides with the saddle one and there is only one critical
point, which is unstable. Thus the Universe can go out of the stable
state and then enter an inflationary era. Therefore, the big bang
singularity can be avoided and a subsequent inflation can occur
naturally.

When $\beta_2\neq 0$, our results show that if the cosmic scale
factor satisfies the condition given in Eq.~(\ref{PG}) and the
equation of state $w$ is larger than a critical value initially, the
Universe can evolve from a stable region to an unstable one with the
decrease of $w$. Therefore, in this case, the big bang singularity
can also be avoided and an inflation can appear naturally. However,
if the cosmic scale factor satisfies the condition given in
Eq.~(\ref{PE}) initially, although the big bang singularity can also
be avoided, the Universe cannot evolve to an inflationary phase with
the evolution of $w$.

Note added: While we are in the stage of revising the manuscript,
the stability of Einstein static universe in a IR modified HL
gravity is discussed in~\cite{Boehmer20093}.

\begin{acknowledgments}

This work was supported in part by the National Natural Science
Foundation of China under Grants No. 10775050, 10705055 and
10935013, the SRFDP under Grant No. 20070542002,  the FANEDD under
Grant No 200922, the Program for NCET (No.09-0144), the National
Basic Research Program of China under Grant No. 2010CB832803, the PCSIRT under Grant
No. IRT0964, and the K.C. Wong Magna Fund in Ningbo
University.

\end{acknowledgments}


\begin{thebibliography}{99}
\bibitem{Ellis2004a}    G. F. R. Ellis and R. Maartens, Class. Quant. Grav. {\bf 21}, 223 (2004).
\bibitem{Ellis2004b}    G. F. R. Ellis, J. Murugan and C. G. Tsagas,  Class. Quant. Grav. {\bf 21}, 233 (2004).
\bibitem{Gibbons1988}  G. W. Gibbons, Nucl. Phys. B {\bf 292}, 784 (1987);
                       Nucl. Phys. B {\bf 310}, 636  (1988).
\bibitem{Carneiro2009} S. Carneiro and R. Tavakol, Phys. Rev. D {\bf 80},  043528 (2009) arXiv: 0907.4795.

\bibitem{Mulryne2005}  D. J. Mulryne, R. Tavakol, J. E. Lidsey and  G. F. R. Ellis,  Phys. Rev. D {\bf 71}, 123512 (2005).
\bibitem{Parisi2007}  L. Parisi, M. Bruni, R. Maartens and K. Vandersloot, Class. Quant. Grav. {\bf 24}, 6243 (2007). 
\bibitem{Wu2009}      P. Wu and H. Yu, JCAP {\bf 05}, 007  (2009) arXiv:0905.3116.
\bibitem{Lidsey2006} J. E. Lidsey and D. J. Mulryne,  Phys. Rev. D {\bf 73}, 083508 (2006).
\bibitem{Bohmer2007} C. G. Boehmer, L. Hollenstein and F. S. N. Lobo,  Phys. Rev. D {\bf 76},  084005 (2007);
                     N. Goheer, R. Goswami and P. K. S. Dunsby, Class. Quant. Grav. {\bf 26}, 105003 (2009) arXiv: 0809.5247;
                     S. del Campo, R. Herrera and P. Labrana, J. Cosmol. Astropart. P. {\bf 0711},  030 (2007);
                     R. Goswami, N. Goheer and P. K. S. Dunsby, Phys. Rev. D {\bf 78},  044011 (2008);
                     U. Debnath, Class. Quant. Grav. {\bf 25},  205019 (2008);
                                        B. C. Paul  and S. Ghose, arXiv: 0809.4131.
\bibitem{Seahra2009} S. S. Seahra and C. G. Bohmer,  Phys. Rev. D {\bf 79}, 064009 (2009).
\bibitem{Bohmer2009} C. G. Boehmer and F. S. N. Lobo,  Phys. Rev. D {\bf 79}, 067504 (2009) arXiv: 0902.2982
\bibitem{Barrow2003}  J. D Barrow, G. Ellis, R. Maartens, C. Tsagas, Class.Quant.Grav. {\bf 20},  L155 (2003).
\bibitem{Clifton2005} T. Clifton, John D. Barrow, Phys.Rev. D {\bf 72},  123003 (2005).
\bibitem{Barrow2009} J. D Barrow, C. G Tsagas, Class. Quant. Grav. {\bf 26},  195003(2009) arXiv:0904.1340.
\bibitem{Boehmer2010} C. G. Boehmer, L. Hollenstein, F. S. N. Lobo,
and  S. S. Seahra, arXiv:1001.1266.
\bibitem{Odrzywolek2009} A. Odrzywolek, Phys. Rev. D {\bf 80}, 103515 (2009).


\bibitem{Horava2009} P. Horava, JHEP {\bf 0903}, 020 (2009) arXiv: 0812.4287;
                    P. Horava, Phys. Rev. D {\bf 79},  084008 (2009) arXiv: 0901.3775;
                    P. Horava, Phys. Rev. Lett. {\bf 102}, 161301 (2009) arXiv: 0902.3657.
\bibitem{Lu2009}  H. Lu, J. Mei and C. N. Pope, Phys. Rev. Lett. {\bf 103}, 091301 (2009) arXiv: 0904. 1595.
\bibitem{Kiritsis2009} E. Kiritsis and G. Kofinas, Nucl. Phys. B {\bf 821}, 467 (2009) arXiv: 0904.1334.
\bibitem{Calcagni2009} G. Calcagni, JHEP {\bf 0909}, 112 (2009) arXiv: 0904.0829.
\bibitem{Mukohyama2009}S. Mukohyama, JCAP {\bf 0906}, 001 (2009) arXiv:0904.2190.
\bibitem{Gao2009}      X.  Gao,  arXiv:0904.4187.
\bibitem{Cai2009}      R. G.  Cai, B. Hu and H. B. Zhang,  Phys. Rev. D {\bf 80},  041501(R) (2009)  arXiv:0905.0255.
\bibitem{Chen2009}  B. Chen, S. Pi and J. Tang, JCAP {\bf 0908}, 007 (2009) arXiv: 0905.2300.
\bibitem{Gao20092}  X. Gao, Y. Wang, R. Brandenberger, A. Riotto, Phys. Rev. D {\bf 81}, 083508 (2010) arXiv: 0905.3821.
\bibitem{Yamamoto2009} K. Yamamoto, T. Kobayashi, G. Nakamura, Phys. Rev. D {\bf 80}, 063514 (2009)  arXiv: 0907.1549.
\bibitem{Wang2009}      A. Wang, R. Maartens, Phys. Rev. D {\bf 81}, 024009 (2010) arXiv: 0907.1748.
\bibitem{Kobayashi2009} T. Kobayashi, Y. Urakawa, M. Yamaguchi, JCAP {\bf 0911}, 015 (2009)  arXiv:0908.1005.
\bibitem{Piao2009}     Y. Piao,  arXiv:0904.4117.
\bibitem{CaiYi2009}    Y. F. Cai, X. Zhang, Phys. Rev.D {\bf 80}, 043520 (2009) arXiv:0906.3341.
\bibitem{Sotiriou2009}T. Sotiriou, M. Visser, and S. Weinfurtner, Phys. Rev. Lett. {\bf 102},
251601 (2009) arXiv:0904.4464; JHEP {\bf 0910}, 033 (2009) arXiv:0905.2798.

\bibitem{Lu20092} U. H. Danielsson, and L. Thorlacius, JHEP, 0903, 070 (2009) arXiv:0812.5088;
 R. G. Cai, L. M. Cao, and N. Ohta, Phys. Rev. D {\bf 80}, 024003 (2009) arXiv:0904.3670;
 R. G. Cai, Y. Liu, and Y. W. Sun, JHEP {\bf 0906}, 010 (2009) arXiv:0904.4104;
 R. G. Cai, L. M. Cao, and N. Ohta, Phys. Lett. B {\bf 679}, 504 (2009) arXiv:0905.0751;
 R. G. Cai and H. Q. Zhang, Phys. Rev. D {\bf 81}, 066003 (2010) arXiv:0911.4867;
 R. G. Cai and N. Ohta, arXiv:0910.2307;
 R. G. Cai and A. Wang, asXiv:1001.0155;
 E. Colgain and H. Yavartanoo, JHEP {\bf 0908}, 021 (2009) arXiv:0904.4357;
 Y. S. Myung and Y.- W. Kim, arXiv:0905.0179;
 Y. S. Myung, Phys. Lett. B {\bf 684}, 158 (2010) arXiv:0908.4132;
 Y. S. Myung, arXiv:0911.0724;
 Y. S. Myung, Phys. Lett. B {\bf 679}, 491 (2009) arXiv:0907.5256;
 Y. S. Myung, Phys. Lett. B {\bf 681}, 81 (2009) arXiv:0909.2075;
 Y. S. Myung, arXiv:0912.3305;
 Y. S. Myung, arXiv:0905.0957;
 Y. S. Myung, Y. W. Kim, and Y.J. Park, arXiv:0910.4428;
 Y. S. Myung, Y. W. Kim, W. S. Son, and Y. J. Park, arXiv:0911.2525;
 A. Kehagias and K. Sfetsos, Phys. Lett. B {\bf 678}, 123 (2009) arXiv:0905.0477;
 R. B. Mann, arXiv:0905.1136;
 S. Chen and J. Jing, Phys. Lett. B {\bf 687}, 124 (2010) arXiv:0905.1409;
 S. Chen and J. Jing, Phys. Rev. D {\bf 80}, 024036 (2009) arXiv:0905.2055;
 D. W. Pang,  arXiv:0905.2678;
 G. Bertoldi, B. A. Burrington, and A. Peet, arXiv:0905.3183;
 M. i. Park, JHEP {\bf 0909}, 123 (2009) arXiv:0905.4480;
 A. Castillo and A. Larranaga, arXiv:0906.4380;
 J. J. Peng and S.Q. Wu, Eur. Phys. J. C {\bf 66}, 325 (2010) arXiv:0906.5121;
 H. W. Lee, Y.W. Kim, and Y.S. Myung, arXiv:0907.3568;
 S. S. Kim, T. Kim, and Y. Kim, arXiv:0907.3093;
 C. Ding, S. Chen, and J. Jing, Phys. Rev. D {\bf 81}, 024028 (2010) arXiv:0909.2490;
 J. Z. Tang and B. Chen, arXiv:0909.4127;
 N. Varghese and V.C. Kuriakose, arXiv:0909.4944;
 D. Y. Chen, H. Yang, and X. T. Zu, arXiv:0910.4821;
 T. Harada, U. Miyamoto, and N. Tsukamoto, 0911.1187;
 D. Capasso, A.P. Polychronakos, arXiv:0911.1535;
 B. R. Majhi, Phys. Lett. B {\bf 686}, 49 (2010) arXiv:0911.3239;
 J. Z. Tang, arXiv:0911.3849;
\bibitem{Lu20093} M. Wang, J. Jing, C. Ding, and S. Chen, Phys. Rev. D {\bf 81}, 083006 (2010) arXiv:0912.4832;
  S. Mukohyama, K. Nakayama, F. Takahashi, and S. Yokoyama, Phys. Lett. B {\bf 679}, 6(2009) arXiv:0905.0055;
 S. Mukohyama, JCAP {\bf 0909}, 005 (2009) arXiv:0906.5069;
 S. K. Rama, Phys. Rev. D {\bf 79}, 124031 (2009) arXiv:0905.0700;
 N. Saridakis, arXiv:0905.3532;
 M. i. Park, JCAP {\bf 1001}, 001 (2010) arXiv:0906.4275;
 S. Koh, arXiv:0907.0850;
 C. Appignani, R. Casadio, and S. Shankaranarayanan, arXiv:0907.3121;
 M. R. Setare, arXiv:0909.0456;
 S. Maeda, S. Mukohyama, and T. Shiromizu, Phys. Rev. D {\bf 80}, 123538 (2009) arXiv:0909.2149;
 S. Carloni, E. Elizalde, and P. J. Silva, arXiv:0909.2219;
 G. Leon, and E. N. Saridakis, JCAP {\bf 0911}, 006 (2009) arXiv:0909.3571;
 B. Chen, S. Pi, and J. Z. Tang, arXiv:0910.0338;
 B. Chen and Q. G. Huang, Phys. Lett. B {\bf 683}, 108 (2010) arXiv:0904.4565;
 S. Dutta and E.N. Saridakis, arXiv:0911.1435;
 I. Bakas, F. Bourliot, D. Lust, and M. Petropoulos, arXiv:0911.2665;
 E. Czuchry, arXiv:0911.3891.
 M. Visser, Phys. Rev. D {\bf 80}, 025011 (2009) arXiv:0902.0590; arXiv:0912.4757;
 P. R. Carvalho and M. Leite, Annals Phys. {\bf 325}, 151 (2010) arXiv:0902.1972;
 A. Volovich and C. Wen, JHEP {\bf 0905}, 087 (2009) arXiv:0903.2455;
 A. Jenkins, Int. J. Mod. Phys. D {\bf 18}, 2249 (2009) arXiv: 0904.0453;
 H. Nikolic, arXiv:0904.3412;
 H. Nastase, arXiv:0904.3604;
 G. E. Volovik, JETP Lett. {\bf  89}, 525 (2009) arXiv:0904.4113;
 D. Orlando and S. Reffert, Class. Quant. Grav. {\bf 26}, 155021 (2009) arXiv:0905.0301;

\bibitem{Lu20094} C. Gao, Phys. Lett. B {\bf 684}, 85 (2010) arXiv:0905.0310;
 T. Nishioka, Class. Quant. Grav. {\bf 26}, 242001 (2009) arXiv:0905.0473;
 A. Ghodsi, arXiv:0905.0836;
 J. B. Jimenez and A. L. Maroto, Phys. Rev. D {\bf 80}, 063512 (2009) arXiv:0905.1245;
 Y. W. Kim, H. W. Lee, and Y. S. Myung, Phys. Lett. B {\bf 682}, 246 (2009) arXiv:0905.3423;
 M. Sakamoto, Phys. Rev. D {\bf 79}, 124038 (2009) arXiv:0905.4326;
 M. Botta-Cantcheff, N. Grandi, and M. Sturla, arXiv:0906.0582;
 Y. S. Myung, Phys. Rev. D {\bf 81}, 064006 (2010) arXiv:0906.0848;
 A. Ghodsi and E. Hatefi, Phys. Rev. D {\bf 81}, 044016 (2010) arXiv:0906.1237;
 A. Kobakhidze, arXiv:0906.5401;
 T. Harko, Z. Kovacs, and F.S. N. Lobo, Phys. Rev. D {\bf 80}, 044021 (2009) arXiv:0907.1449;
 I. Adam, I. V. Melnikov and S. Theisen, JHEP {\bf 0909}, 130 (2009) arXiv:0907.2156;
 N. Afshordi, Phys. Rev. D {\bf 80}, 081502 (2009) arXiv:0907.5201;
 I. Cho and G. Kang, arXiv:0909.3065;
 T. Suyama, JHEP {\bf 01}, 093 (2010) arXiv:0909.4833;
 D. Capasso and A.P.Polychronakos, JHEP {\bf 1002}, 068 (2010) arXiv:0909.5405;
 S. K. Rama, arXiv:0910.0411;
 D. Momeni, arXiv:0910.0594;
 M. Park, arXiv:0910.1917; arXiv:0910.5117;
 D. Benedetti and J. Henson, Phys. Rev. D {\bf 80}, 124036 (2009) arXiv:0911.0401;
 Q. Exirifard, arXiv:0911.4343;
 W. Chao, arXiv:0911.4709;
 R. Garattini, arXiv:0912.0136;
 G. Calcagni, Phys. Rev. D {\bf 81}, 044006 (2010) arXiv:0905.3740;
 C. Bogdanos, and E.N. Saridakis,Class. Quant. Grav. {\bf 27}, 075005 (2010) arXiv:0907.1636;
 J. Kluson, arXiv:0904.1314; arXiv:0910.5852;
 J. Chen, and Y. Wang, Int. J. Mod. Phys. A {\bf 25},  1439 (2010) arXiv:0905.2786;
 S. Nojiri and S.D. Odintsov, Phys. Rev. D {\bf 81}, 043001 (2010) arXiv:0905.4213;
 C. Germani, A. Kehagias, and K. Sfetsos, arXiv:0906.1201.
 D. Blas, O.Pujolas, and S. Sibiryakov, arXiv:0909.3525; arXiv:0912.0550.
 T. Takahashi and J. Soda, Phys. Rev. Lett. {\bf 102}, 231301 (2009) arXiv:0904.0554.
 R. Brandenberger, arXiv:0904.2835;
 Y. Cai, E. N. Saridakis, JCAP {\bf0910}, 020  (2009) arXiv:0906.1789.
 S. Mukohyama,  arXiv:0905.3563.
 A. Papazoglou and T. P. Sotiriou, Phys. Lett. B {\bf 685}, 197 (2010) arXiv:0911.1299.

\bibitem{Lu20095}   K. Izumi and S. Mukohyama, Phys. Rev. D {\bf 81},  044008 (2010) arXiv:0911.1814;
 R. Iengo, J. G. Russo, and M. Serone, JHEP {\bf 11},  020  (2009) arXiv:0906.3477;
 R. A. Konoplya, Phys. Lett. B {\bf 679}, 499  (2009) arXiv:0905.1523;
 T. Harko, Z. Kovacs and F. S. N. Lobo, arXiv:0908.2874;
 L. Iorio and M. L. Ruggiero, arXiv:0909.2562; arXiv:0909.5355;
 G. Amelino-Camelia, L. Gualtieri, and F. Mercati, Phys. Lett. B {\bf 686},  283(2010) arXiv:0911.5360;
 A. Wang and Y. Wu, JCAP, 07, 012 (2009)arXiv: 0905.4117;
 J. Greenwald, A. Papazoglou, A. Wang,  arXiv:0912.0011;
 M. R. Setare, M. Jamil,  arXiv:1001.1251;
 J. Jing, L. Wang, S. Chen,  arXiv:1001.1472;
 Q. Cao, Y. Chen, K. Shao,  arXiv:1001.2597;
 F. S. N. Lobo, T. Harko, Z.  Kovacs, arXiv:1001.3517;
 M. R. Setare, D. Momeni,  arXiv:1001.3767;
 Y. Myung, Y. Kim, W. Son, Y. Park, arXiv:1001.3921;
 M. Chaichian, S. Nojiri, S. D. Odintsov, M. Oksanen, A. Tureanu,  arXiv:1001.4102;
 T. Jacobson,  arXiv:1001.4823.

\bibitem{Setare2009} M. R. Setare, D. Momeni, arXiv:0911.1877.
\bibitem{Minamitsuji2010} M. Minamitsuji, Phys. Lett. B {\bf 684}, 194 (2010).
\bibitem{Boehmer20093}C. G. Boehmer, F. S. N. Lobo, arXiv:0909.3986.


\end{thebibliography}
\end{document}